\documentclass[onecolumn]{IEEEtran}
\usepackage{multicol}
\usepackage{times}
\usepackage{times,amssymb,amsmath,amsfonts,float,color,cite,bbm,mathrsfs,float,stmaryrd}
\usepackage{enumerate,bm}
\usepackage{amsfonts}
\usepackage{latexsym}
\usepackage{theorem}
\usepackage{graphicx}
\usepackage{subcaption}
\usepackage{enumitem}
\usepackage{algorithm,algpseudocode}
\usepackage[normalem]{ulem}
\usepackage{hyperref}
\usepackage{cleveref}

\usepackage{algorithm,algpseudocode}

\makeatletter
\newcommand{\removelatexerror}{\let\@latex@error\@gobble}
\makeatletter
\newcommand{\proofpart}[2]{%
	\par
	\addvspace{\medskipamount}%
	\noindent\emph{Part #1: #2}\par\nobreak
	\addvspace{\smallskipamount}%
	\@afterheading
}

\newtheorem{theorem}{Theorem}[section]
\newtheorem{lemma}[theorem]{Lemma}

\newtheorem{proposition}[theorem]{Proposition}

\newtheorem{corollary}[theorem]{Corollary}
\newtheorem{definition}[theorem]{Definition}

\newtheorem{example}[theorem]{Example}
\newtheorem{remark}[theorem]{Remark}
\newtheorem{construction}[theorem]{Construction}

\renewcommand{\thefigure}{\thesection.\arabic{figure}}
\usepackage{amsmath}

\allowdisplaybreaks

\makeatletter
\renewcommand{\@endtheorem}{\endtrivlist}
\makeatother

\newcommand\remove[1]{}

\makeatletter
\renewcommand{\thefigure}{{\@arabic\c@figure}}
\renewcommand{\fnum@figure}{{\bf Figure\,\thefigure}}
\makeatother

\newcommand\nc\newcommand
\nc{\cA}{\mathcal{A}}\nc{\cB}{\mathcal{B}}\nc{\cC}{\mathcal{C}}\nc{\cD}{\mathcal{D}}
\nc{\cE}{\mathcal{E}}\nc{\cF}{\mathcal{F}}\nc{\cG}{\mathcal{G}}\nc{\cH}{\mathcal{H}}
\nc{\cI}{\mathcal{I}}\nc{\cJ}{\mathcal{J}}\nc{\cK}{\mathcal{K}}\nc{\cL}{\mathcal{L}}
\nc{\cM}{\mathcal{M}}\nc{\cN}{\mathcal{N}}\nc{\cO}{\mathcal{O}}\nc{\cP}{\mathcal{P}}
\nc{\cQ}{\mathcal{Q}}\nc{\cR}{\mathcal{R}}\nc{\cS}{\mathcal{S}}\nc{\cT}{\mathcal{T}}
\nc{\cU}{\mathcal{U}}\nc{\cV}{\mathcal{V}}\nc{\cW}{\mathcal{W}}\nc{\cX}{\mathcal{X}}
\nc{\cY}{\mathcal{Y}}\nc{\cZ}{\mathcal{Z}}

\nc{\bba}{\mathbf{a}}\nc{\bbb}{\mathbf{b}}\nc{\bbc}{\mathbf{c}}\nc{\bbd}{\mathbf{d}}
\nc{\bbe}{\mathbf{e}}\nc{\bbf}{\mathbf{f}}\nc{\bbg}{\mathbf{g}}\nc{\bbh}{\mathbf{h}}
\nc{\bbi}{\mathbf{i}}\nc{\bbj}{\mathbf{j}}\nc{\bbk}{\mathbf{k}}\nc{\bbl}{\mathbf{l}}
\nc{\bbm}{\mathbf{m}}\nc{\bbn}{\mathbf{n}}\nc{\bbo}{\mathbf{o}}\nc{\bbp}{\mathbf{p}}
\nc{\bbq}{\mathbf{q}}\nc{\bbr}{\mathbf{r}}\nc{\bbs}{\mathbf{s}}\nc{\bbt}{\mathbf{t}}
\nc{\bbu}{\mathbf{u}}\nc{\bbv}{\bm{v}}\nc{\bbw}{\mathbf{w}}\nc{\bfx}{\mathbf{x}}
\nc{\bby}{\mathbf{y}}\nc{\bbz}{\mathbf{z}}

\nc{\bbA}{\mathbf{A}}\nc{\bbB}{\mathbf{B}}\nc{\bbC}{\mathbf{C}}\nc{\bbD}{\mathbf{D}}
\nc{\bbE}{\mathbf{E}}\nc{\bbF}{\mathbf{F}}\nc{\bbG}{\mathbf{G}}\nc{\bbH}{\mathbf{H}}
\nc{\bbI}{\mathbf{I}}\nc{\bbJ}{\mathbf{J}}\nc{\bbK}{\mathbf{K}}\nc{\bbL}{\mathbf{L}}
\nc{\bbM}{\mathbf{M}}\nc{\bbN}{\mathbf{N}}\nc{\bbO}{\mathbf{O}}\nc{\bbP}{\mathbf{P}}
\nc{\bbQ}{\mathbf{Q}}\nc{\bbR}{\mathbf{R}}\nc{\bbS}{\mathbf{S}}\nc{\bbT}{\mathbf{T}}
\nc{\bbU}{\mathbf{U}}\nc{\bbV}{\mathbf{V}}\nc{\bbW}{\mathbf{W}}\nc{\bfX}{\mathbf{X}}
\nc{\bbY}{\mathbf{Y}}\nc{\bbZ}{\mathbf{Z}}

\nc{\sA}{\mathsf{A}}\nc{\sB}{\mathsf{B}}\nc{\sC}{\mathsf{C}}\nc{\sD}{\mathsf{D}}
\nc{\sE}{\mathsf{E}}\nc{\sF}{\mathsf{F}}\nc{\sG}{\mathsf{G}}\nc{\sH}{\mathsf{H}}
\nc{\sI}{\mathsf{I}}\nc{\sJ}{\mathsf{J}}\nc{\sK}{\mathsf{K}}\nc{\sL}{\mathsf{L}}
\nc{\sM}{\mathsf{M}}\nc{\sN}{\mathsf{N}}\nc{\sO}{\mathsf{O}}\nc{\sP}{\mathsf{P}}
\nc{\sQ}{\mathsf{Q}}\nc{\sR}{\mathsf{R}}\nc{\sS}{\mathsf{S}}\nc{\sT}{\mathsf{T}}
\nc{\sU}{\mathsf{U}}\nc{\sV}{\mathsf{V}}\nc{\sW}{\mathsf{W}}\nc{\sX}{\mathsf{X}}
\nc{\sY}{\mathsf{Y}}\nc{\sZ}{\mathsf{Z}}

\newcommand{\mathset}[1]{\left\{#1\right\}}
\newcommand{\abs}[1]{\left|#1\right|}

\newcommand{\floorenv}[1]{\left\lfloor #1 \right\rfloor}
\newcommand{\parenv}[1]{\left( #1 \right)}

\nc{\set}[1]{\llbracket #1 \rrbracket}

\newcommand{\bal}[1]{\begin{align}\label{#1}}
\newcommand{\eal}{\end{align}}

\renewcommand{\le}{\leqslant}
\renewcommand{\leq}{\leqslant}
\renewcommand{\ge}{\geqslant}
\renewcommand{\geq}{\geqslant}

\renewcommand{\Bbb}{\mathbb}

\newcommand{\Lref}[1]{Lem\-ma\,\ref{#1}}

\renewcommand{\Bbb}{\mathbb}

\newcommand{\N}{{\Bbb N}}

\newcommand{\bbss}{\bbs^{\star}}

\newcommand{\Rnum}[1]{\lowercase\expandafter{\romannumeral #1\relax}}
\newcommand{\RNum}[1]{\uppercase\expandafter{\romannumeral #1\relax}}

\newcommand{\ouc}[1]{#1^{\text{\textcentoldstyle}}}

\newcommand{\0}{\mathbf{0}}
\newcommand{\1}{\mathbf{1}}

\nc{\vt}{\vartheta}
\nc{\vtp}{\vartheta_{\bbP}}
\nc{\vtpk}{\vartheta_{\bbP,k}}

\newcommand{\wth}{\mathsf{wt}_H\! }
\DeclareMathOperator{\swap}{Swap}

\outer\def\proclaim #1. #2\par{\medbreak
 \noindent{\bf#1.\enspace}{\sl#2\par}%
 \ifdim\lastskip<\medskipamount \removelastskip\penalty55\medskip\fi}

\begin{document}

\title{{Reconstruction of a Single String from a Part of its Composition Multiset}}

\author{\IEEEauthorblockN{Zuo~Ye and Ohad ~Elishco}
}

\maketitle

{\renewcommand{\thefootnote}{}\footnotetext{

\vspace{-.2in}
 
\noindent\rule{1.5in}{.4pt}

{ The authors are with the School of Electrical and Computer Engineering, Ben-Gurion University of the Negev, 
Beer Sheva, Israel. Email: \{zuoy,ohadeli\}@bgu.ac.il. 
}
}}



\begin{abstract}
Motivated by applications in polymer-based data storage, we study the problem of reconstructing a string from part of its composition multiset.
We give a full description of the structure of the strings that cannot be uniquely reconstructed (up to reversal) from their multiset of all of their prefix-suffix compositions. Leveraging this description, we prove that for all $n\ge 6$, there exists a string of length $n$ that cannot be uniquely reconstructed up to reversal. Moreover, for all $n\ge 6$, we explicitly construct the set consisting of all length $n$ strings that can be uniquely reconstructed up to reversal. As a by product, we obtain that any binary string can be constructed using Dyck strings and Catalan-Bertrand strings. 

For any given string $\bbs$, we provide a method to explicitly construct the set of all strings with the same prefix-suffix composition multiset as $\bbs$, as well as a formula for the size of this set. 
As an application, we construct a composition code of maximal size. 
Furthermore, we construct two classes of composition codes which can respectively correct composition missing errors and mass reducing substitution errors.

In addition, we raise two new problems: reconstructing a string from its composition multiset when at most a constant number of substring compositions are lost; reconstructing a string when only given its compositions of substrings of length at most $r$. For each of these setups, we give suitable codes under some conditions.
\end{abstract}


\section{Introduction}
In this information age, transmission, processing, and storage of data are playing an increasingly important role. As human life becomes more and more dependent on data, exponentially increasing amounts of data are generated every year. According to the IDC, the global datasphere will grow from 33 zettabytes in 2018 to 175 zettabytes by 2025 \cite{IDC2018}. Therefore, the requirements for the density and longevity of storage devices are higher and higher. 
However, existing data storage media are not particularly advantageous in terms of these two requirements. And this is a major stimulus for the search for storage paradigms that offer storage densities at the nanoscale. Recently, several molecular storage techniques, which can provide significantly higher storage densities and longevities, have been proposed \cite{science2012,nature2013,wz4,wz5,wz6,wz7,wz8,wz9,wz10,wz11,wz12,wz13,wz14,wz15,wz16}. DNA-based data storage is one of such promising data storage media, but the systems come with slow and error-prone read/write platforms, and have several constraints including a prohibitively expensive synthesis and sequencing process.

Polymer-based data storage systems \cite{wz7,wz16}, as an alternative to DNA-based data storage systems, are emerging as the next-generation data storage medium. Comparing to DNA-based data storage systems, they offer high storage densities at lower cost and lower readout latency. In such platforms, two molecules (phosphodiesters) with significantly different masses are synthesized, one of which represents the bit $0$ while the other represents the bit $1$. In the write-in process, a string of binary information is encoded into a string (poly-phosphodiester) of these two molecules which are stitched together by means of phosphate bonds. In the readout process, a tandem mass (MS/MS) spectrometer is used to break some of these phosphate bonds, thereby creating fragments of various lengths, and then output the masses of the resulting fragments. Then the information of masses is utilized to reconstruct the whole string.

There have been several works on this topic. The first one is \cite{wz1}, the results of which are obtained under the following two assumptions.
\begin{enumerate}
  \item[\textit{A}1:] One can uniquely infer the composition, i.e., the number of $0$'s and $1$'s, of a fragment string from its mass.
  \item[\textit{A}2:] When a polymer is broken down for mass spectrometry analysis, the masses of all its fragments will be observed with identical frequency.
\end{enumerate}

The authors of \cite{wz1} proved that when $n\le 7$, or $n$ is one less than a prime or one less than twice a prime, any string of length $n$ can be uniquely reconstructed up to reversal from its composition multiset, i.e., the multiset consisting of the compositions of all its substrings (fragments). 
Later, under the same assumptions, the authors of \cite{wz3} viewed the problem from a coding-theoretic perspective and demonstrated that: 1) one can add at most logarithmic redundancy to an information string so that unique reconstruction is possible, even if the information string itself cannot be uniquely reconstructed up to reversal; 2) one can add at most logarithmic redundancy to an information string so that unique reconstruction is possible, even in the presence of errors. Following \cite{wz3}, in \cite{wz17} the authors investigated more types of errors and constructed corresponding codes. 
Later, a new algorithm for the problem of string reconstruction from its composition multiset was introduced \cite{wz18}. In case of no backtracking, this new algorithm has a time complexity $O(n^2)$, which is lower by a $\log_2(n)$ factor than that of the algorithm proposed in \cite{wz1}, where $n$ is the length of the string. Furthermore, the authors constructed a set of strings that can be uniquely reconstructed, the size of which is larger by a linear factor than that of the code given in \cite{wz3}.

Recall that all of these works are based on \textit{Assumption} 2. However, it is often not practical to assume that the masses of all fragments can be observed, since breaking the string in one rather than two positions is easier to perform. In \cite{wz2}, the authors studied the problem of reconstructing strings from the multiset consisting of the compositions of all their prefixes and suffixes, rather than focusing on the reconstruction of a single string, they investigated a more general problem: reconstruction of a set of at most $h$ strings from the union of their prefix-suffix composition multisets, where $h$ is a given positive integer. By using $B_h$ codes and Dyck strings, they explicitly constructed a code $C(n,h)$, in which for any subset of $C(n,h)$ of size at most $h$, one can uniquely reconstruct this subset, where $n$ is the length of the strings and is of the form $m+\frac{17}{2}\sqrt{m}$ with $m$ being an positive integer. Furthermore, they studied two types of error models, and based on $C(n,h)$, they constructed several codes that can correct such errors.

Obviously, if no error occurs in the (prefix-suffix) composition multiset, all the works mentioned above can be viewed as special cases of a more general problem: how to reconstruct strings from a part of their composition multisets? In this paper, we continue this line of research and we study three subclasses of this problem. The first one is how to reconstruct a single string from its prefix-suffix composition multiset.  Intuitively, one can deem that this is a critical step to go further and fully understand the more general setup considered in \cite{wz2}. Therefore we only concentrate on the reconstruction of a single string and ask the following questions.
\begin{description}
  \item[Question 1:] For which $n$, all strings of length $n$ can be uniquely reconstructed up to reversal?
  \item[Question 2:] Given $n$, can we construct all strings of length $n$ that cannot be uniquely reconstructed up to reversal in a uniform way? What is the number of such strings?
  \item[Question 3:] Given $n$, can we construct all the strings of length $n$ that can be uniquely reconstructed up to reversal in a uniform way? What is the number of such strings?
  \item[Question 4:] Given a string $\bbs$, how can we find all the strings that share a common prefix-suffix composition multiset with $\bbs$? What is the number of such strings?
  \item[Question 5:] For the specific case of $h=1$, can we improve upon the construction of $C(n,1)$, either in the error-free case or in the erroneous case?
\end{description}

In this paper, we will completely solve these questions. Our basic method is the generating polynomial method first introduced in \cite{wz1}. 
In addition, we find that all binary strings can be constructed by Dyck strings and Catlan-Bertrand strings (see Corollary \ref{cor_all}).

Secondly, we turn our attention to the second subclass of the general problem: how to reconstruct a string from its composition multiset if at most a constant number of compositions were lost in the reading process. The third subclass of the general problem is how to reconstruct a string if we are only given its compositions of substrings of length at most $r$. Both setups are new and well motivated (a version of the second setup was studied in \cite{wz17}), and we provide codes suitable for each of these new setups.

The paper is organized as follows. \Cref{sec_preliminary} is devoted to introducing the problem, the relevant terminologies and notations. 
In \Cref{sec_notunique}, we study strings that cannot be uniquely reconstructed up to reversal and give a uniform way to construct them. 
Thus, Question 1 and the first part of Question 2 are answered. 
In \Cref{sec_unique}, we study strings that can be uniquely reconstructed up to reversal and give a uniform way to construct them. 
In this section, the second part of Question 2 and Question 3 are answered. 
In \Cref{sec_equiv}, we present a method to construct and enumerate all strings that share a common prefix-suffix composition multiset with a given string. 
Thus, answering Question 4. 
Based on the results in previous sections, in \Cref{sec_codes} we construct several codes which improve upon the best known codes. 
We also construct codes that can correct errors of the type considered in \cite{wz2}. 
This answers Question 5 affirmatively. 
In \Cref{sec_delcomp} we study the reconstruction problem from the entire composition multiset. 
We study the model in which some compositions are deleted from the composition multiset, and present a code construction that can correct up to $4t$ asymmetric errors, where $t\ge 2$ is a constant.
In Section \ref{sec:RLL} we present a new model called the $r$-length limited composition multiset model, and provide a simple coding scheme for the case $r$ is a linear function of 
the length of the words $n$. 
In \Cref{sec_conclusion}, we conclude this paper and discuss future problems.

\section{Preliminaries}\label{sec_preliminary}
Let $\Sigma_2\triangleq\{0,1\}$ be the binary alphabet. For any string $\bbs\in\Sigma_2^n$, denote by $c(\bbs)=0^{n-w}1^{w}$ the \textit{composition} of $\bbs$, where $w=wt_{H}(\bbs)$ is the Hamming weight of $\bbs$. If $\bbs=s_1\cdots s_n$, we denote $\bbs_{i}^{j}=s_i\cdots s_j$ for any $1\le i\le j\le n$. 
If $j\leq i-1$, we define $\bbs_{i}^{j}\triangleq\emptyset$. In particular, $\bbs_1^{j}$ and $\bbs_i^{n}$ are called the prefix and suffix of $\bbs$ respectively. 
\begin{definition}
For a string $\bbs\in\Sigma_2^n$, we denote by $\cM(\bbs)$ the multiset of compositions of all prefixes and suffixes of $\bbs$, i.e., 
\[\cM(\bbs):=\mathset{([1,j],c(\bbs_1^j)) ~:~ 1\leq j\leq n}\cup \Big\{([i,n],c(\bbs_i^n)) ~:~ 1\leq i\leq n\Big\}.\] 
By abuse of notation, we sometimes refer only to the right arguments in $\cM(\bbs)$.
We call $\cM(\bbs)$ the \textit{prefix-suffix composition multiset} of $\bbs$. 
\end{definition}
For example, if $\bbs=01010$, then
\[
\mathcal{M}(\bbs)=\left\{0,0,01,01,0^21,0^21,0^21^2,0^21^2,0^31^2\right\}.
\]

For any $\bbs=s_1\cdots s_n$, we denote $\bbss=s_ns_{n-1}\cdots s_1$ and call it the \textit{reversal} of $\bbs$. It is clear that $\mathcal{M}(\bbs)=\mathcal{M}(\bbss)$ for any string $\bbs$. We say a string $\bbs$ can be \textit{uniquely reconstructed up to reversal} if there is no string $\bbt$ other than $\bbs$ and $\bbss$, such that $\mathcal{M}(\bbt)=\mathcal{M}(\bbs)$.

With a given string $\bbs=s_1\cdots s_n\in\Sigma_2^n$, we can associate a bivariate polynomial $P_{\bbs}(x,y)$ with coefficients in $\{0,1\}$. 
In this paper, for a given bivariate polynomial $P(x,y)$, we always let $\left(P(x,y)\right)_i$ denote the term of total degree $i$ of $P(x,y)$. 
The next polynomial was firstly introduced in \cite{wz1}. 
\begin{definition}
Let $\bbs\in\Sigma_2^n$ be a string. Define the \textit{generating polynomial} of $\bbs$, denoted $P_{\bbs}(x,y)$, as $P_{\bbs}(x,y)_0\triangleq 1$ 
and for any $1\le i\le n$, 
\[P_{\bbs}(x,y)_i\triangleq
\begin{cases}
  y\left(P_{\bbs}(x,y)\right)_{i-1}, & \mbox{if } s_i=0, \\
  x\left(P_{\bbs}(x,y)\right)_{i-1}, & \mbox{if } s_i=1.
\end{cases}
\]
\end{definition}
For example, if $\bbs=01010$, then $P_{\bbs}(x,y)=1+y+xy+xy^2+x^2y^2+x^2y^3$. 
By definition, we can also write $P_{\bbs}(x,y)$ as
\begin{equation}\label{eq_generating}
P_{\bbs}(x,y)=\mathop{\sum}\limits_{i=0}^{n}x^{a_i}y^{i-a_i},
\end{equation}
where $a_0\triangleq0$ and $a_i=\wth(\bbs_1^{i})$ for any $1\le i\le n$. The \textit{reciprocal} of $P_{\bbs}(x,y)$ is defined to be the polynomial $P_{\bbs}^{*}(x,y)=x^{a_n}y^{n-a_n}P_{\bbs}(\frac{1}{x},\frac{1}{y})$. Thus we have
\begin{equation}\label{eq_dualgenerating}
P_{\bbs}^{*}(x,y) =\mathop{\sum}\limits_{i=0}^{n}x^{w-a_{n-i}}y^{i-(w-a_{n-i})},
\end{equation}
where $w\triangleq a_n$.

The relationship between $P_{\bbs}(x,y)$ and $P_{\bbss}(x,y)$ is described in the following lemma.
\begin{lemma}\cite[Section 7]{wz1}\label{lem_genrev}
For any string $\bbs$, we have $P_{\bbss}(x,y)=P_{\bbs}^{*}(x,y)$.
\end{lemma}

\begin{remark}\label{rmk_weights}
    It is immediate to verify that $\parenv{P_{\bbs}(x,y)}_i=\parenv{P_{\bbs}^{*}(x,y)}_i$ if and only if $\parenv{P_{\bbs}(x,y)}_{n-i}=\parenv{P_{\bbs}^{*}(x,y)}_{n-i}$ for every $0\leq i\leq \floorenv{\frac{n}{2}}$ and $\bbs\in\Sigma^n_2$. 
    By Lemma \ref{lem_genrev} and the relationship between a generating polynomial and its corresponding string, this is equivalent to 
    $\wth \parenv{\bbs_{1}^{i}}=\wth\parenv{(\bbss)_{1}^{i}}$ if and only if $\wth\parenv{\bbs_{1}^{n-i}}=\wth\parenv{(\bbss)_{1}^{n-i}}$. 
    This, in turn, is equivalent to $\wth\parenv{\bbs_{1}^{i}}=\wth\parenv{\bbs_{n-i+1}^{n}}$ if and only if $\wth\parenv{\bbs_{1}^{n-i}}=\wth\parenv{\bbs_{i+1}^{n}}$.
\end{remark}

The following lemma, which can be easily verified, is key to our purpose.
\begin{lemma}\label{lem_confusion}
  For any two strings $\bbs$ and $\bbt$, $\mathcal{M}(\bbs)=\mathcal{M}(\bbt)$ if and only if $P_{\bbs}(x,y)+P_{\bbss}(x,y)=P_{\bbt}(x,y)+P_{\bbt^{*}}(x,y)$, i.e., $P_{\bbs}(x,y)+P_{\bbs}^{*}(x,y)=P_{\bbt}(x,y)+P_{\bbt}^{*}(x,y)$.
\end{lemma}
In the following section we use Lemma \ref{lem_confusion} to identify strings that are not uniquely reconstructible up to reversal.

\section{Strings that cannot be uniquely reconstructed up to reversal}\label{sec_notunique}

Let $\bbs\in\Sigma_2^n$ be a string of length $n$. By Lemma \ref{lem_confusion}, to find a string $\bbt$ such that $\mathcal{M}(\bbs)=\mathcal{M}(\bbt)$, we need to find a  generating polynomial $Q(x,y)$ for some string $\bbt$ with coefficients in $\{0,1\}$ such that
\begin{equation}\label{eq_confusion}
P_{\bbs}(x,y)+P_{\bbs}^{*}(x,y)=Q(x,y)+Q^{*}(x,y).
\end{equation}

Recall that the coefficients of $P_{\bbs}(x,y)$ and $Q(x,y)$ are $0$ or $1$. Therefore, if \Cref{eq_confusion} holds, then either $\left(Q(x,y)\right)_{i}=\left(P_{\bbs}(x,y)\right)_{i}$ or $\left(Q(x,y)\right)_{i}=\left(P_{\bbs}^{*}(x,y)\right)_{i}$. In other words, the polynomial $Q(x,y)$ must be obtained by replacing some $\left(P_{\bbs}(x,y)\right)_{i}$'s with the corresponding $\left(P_{\bbs}^{*}(x,y)\right)_{i}$'s, or by replacing some $\left(P_{\bbs}^{*}(x,y)\right)_{i}$'s with the corresponding $\left(P_{\bbs}(x,y)\right)_{i}$'s. On the other hand, if $Q(x,y)$ is obtained in this way, \Cref{eq_confusion} clearly holds.

Of course, it is meaningless to swap $\left(P_{\bbs}(x,y)\right)_{i}$ and $\left(P_{\bbs}^{*}(x,y)\right)_{i}$ if $\left(P_{\bbs}(x,y)\right)_{i}=\left(P_{\bbs}^{*}(x,y)\right)_{i}$. Besides, it always holds that $\left(P_{\bbs}(x,y)\right)_{n}=\left(P_{\bbs}^{*}(x,y)\right)_{n}$. So we define the set
$$
D\triangleq\mathset{i~:~1\le i< n,\left(P_{\bbs}(x,y)\right)_{i}\ne\left(P_{\bbs}^{*}(x,y)\right)_{i}}.
$$
By Remark \ref{rmk_weights}, $i\in D$ if and only if $n-i\in D$.
If $D=\emptyset$, then $\left(P_{\bbs}(x,y)\right)_{i}=\left(P_{\bbs}^{*}(x,y)\right)_{i}$ for all $1\le i\le n$, which is equivalent to $\bbs=\bbss$ and thus $\bbs$ can be uniquely reconstructed up to reversal. So for our aim, we assume $D\ne\emptyset$. For a nonempty subset $S$ of $D$, let $Q_{S,1}(x,y)$ be obtained by replacing all the $\left(P_{\bbs}(x,y)\right)_{j}$'s with $\left(P_{\bbs}^{*}(x,y)\right)_{j}$'s, and let $Q_{S,2}(x,y)$ be obtained by replacing all the $\left(P_{\bbs}^{*}(x,y)\right)_{j}$'s with $\left(P_{\bbs}(x,y)\right)_{j}$'s, where $j\in S$. If $S$ is clear from the context, we can omit $S$ in the subscript.
\begin{example}\label{examp_dual}
Let $\bbs=1000$. Then $P_{\bbs}(x,y)=1+x+xy+xy^2+xy^3$ and $P_{\bbs}^{*}(x,y)=1+y+y^2+y^3+xy^3$. So $D=\mathset{1,2,3}$. Let $S=\mathset{1}$ and $T=\mathset{1,3}$. Then
$$
\begin{array}{l}
Q_{S,1}(x,y)=1+y+xy+xy^2+xy^3,\\
Q_{S,2}(x,y)=1+x+y^2+y^3+xy^3,\\
Q_{T,1}(x,y)=1+y+xy+y^3+xy^3,\\
Q_{T,2}(x,y)=1+x+y^2+xy^2+xy^3.
\end{array}
$$
Clearly, $Q_{S,1}(x,y)\ne Q_{S,2}^{*}(x,y)$ and $Q_{T,1}(x,y)= Q_{T,2}^{*}(x,y)$.
\end{example}
Example \ref{examp_dual} tells that we should carefully choose $S$ to ensure $Q_{S,1}(x,y)= Q_{S,2}^{*}(x,y)$.
\begin{lemma}\label{lem_dual}
$Q_{S,1}(x,y)= Q_{S,2}^{*}(x,y)$ if and only if $S$ satisfies the condition that $j\in S$ if and only if $n-j\in S$.
\end{lemma}
\begin{IEEEproof}
The $\Leftarrow$ direction is easy to verify be the definition of $Q_{S,1}(x,y)$ and $Q_{S,2}(x,y)$.
Suppose that $P_{\bbs}(x,y)$ is given in \Cref{eq_generating}. Then $P_{\bbs}^{*}(x,y)$ is given in \Cref{eq_dualgenerating}.
  We prove the $\Rightarrow$ direction by contradiction.
  
  Suppose on the contrary that there is a $j\in S$ such that $n-j\notin S$. By the definition of $Q_{S,1}(x,y)$, we can assume
   $$
   Q_{S,1}(x,y)=\mathop{\sum}\limits_{i=0}^{n}x^{d_i}y^{i-d_i}.
   $$
   Then 
   $$
   Q_{S,1}^{*}(x,y) =\mathop{\sum}\limits_{i=0}^{n}x^{w-d_{n-i}}y^{i-(w-d_{n-i})}.
   $$
   In particular, $\parenv{Q_{S,1}^{*}(x,y)}_{j}=x^{w-d_{n-j}}y^{j-(w-d_{n-j})}$. Since $n-j\notin S$, we have $d_{n-j}=a_{n-j}$. Comparing with \Cref{eq_dualgenerating}, we conclude that $\parenv{Q_{S,1}^{*}(x,y)}_{j}=\parenv{P_{\bbs}^{*}(x,y)}_{j}$. On the other hand, we have $\parenv{Q_{S,2}(x,y)}_{j}=\parenv{P_{\bbs}(x,y)}_{j}$ and $\parenv{P_{\bbs}(x,y)}_{j}\ne\parenv{P_{\bbs}^{*}(x,y)}_{j}$ since $j\in S$. Therefore, $\parenv{Q_{S,1}^{*}(x,y)}_{j}\ne\parenv{Q_{S,2}(x,y)}_{j}$, which is a contradiction.
\end{IEEEproof}

By \Lref{lem_dual}, to find a desired $Q(x,y)$ in \Cref{eq_confusion}, it is enough to consider subsets $D\cap\mathset{1,\ldots,\floorenv{\frac{n}{2}}}$. So in the rest of this section, we always assume that $S$ is a subset of $D\cap\mathset{1,\ldots,\floorenv{\frac{n}{2}}}$ and denote $Q_{S,1}(x,y)=Q_{S\cup(n-S),1}(x,y)$ and $Q_{S,2}(x,y)=Q_{S\cup(n-S),2}(x,y)$, where $n-S=\mathset{n-j~:~j\in S}$.

\subsection{A Special Case}
In this subsection, we study the special case when $\abs{S}=1$. On the one hand, this special case inspires our discussion of the general case in next subsection. On the other hand, in this special case, we can obtain more accurate descriptions on the structures of strings.

Let $P_{\bbs}(x,y)$ and $P_{\bbs}^{*}(x,y)$ be given in \Cref{eq_generating} and \Cref{eq_dualgenerating} respectively.
Firstly, We should describe on which terms of $P_{\bbs}(x,y)$ and $P_{\bbs}^{*}(x,y)$ the ``swapping" operation can be performed, such that the obtained polynomial $Q(x,y)$ and is indeed a generating polynomial of some string. 
Let $Q_1(x,y)$ be obtained by replacing $\left(P_{\bbs}(x,y)\right)_{j}$ with $\left(P_{\bbs}^{*}(x,y)\right)_{j}$ and $Q_2(x,y)$ be obtained by replacing $\left(P_{\bbs}^{*}(x,y)\right)_{j}$ with $\left(P_{\bbs}(x,y)\right)_{j}$, where $1\le j\le n-1$. Then $P_{\bbs}(x,y)+P_{\bbs}^{*}(x,y)=Q_1(x,y)+Q_2(x,y)$. To make sure that $Q_1(x,y)$ and $Q_2(x,y)$ are both generating polynomials, the following inequalities must hold:
\begin{equation*}
  \left\{
  \begin{array}{l}
    w-a_{n-j+1}\le a_j\le w-a_{n-j-1}\\
    j-1-(w-a_{n-j+1})\le j-a_j\le j+1-(w-a_{n-j-1})\\
    a_{j-1}\le w-a_{n-j}\le a_{j+1}\\
    j-1-a_{j-1}\le j-(w-a_{n-j})\le j+1-a_{j+1}
  \end{array}
  \right.
\end{equation*}
Simplifying these inequalities, we get
\begin{equation}\label{eq_swap}
  \left\{
  \begin{array}{l}
    w-a_{n-j+1}\le a_j\le w-a_{n-j-1}\\
    w-a_{n-j-1}-1\le a_j\le w-a_{n-j+1}+1\\
    w-a_{j+1}\le a_{n-j}\le w-a_{j-1}\\
   w-a_{j-1}-1\le a_{n-j}\le w-a_{j+1}+1
  \end{array}
  \right.
\end{equation}
If \Cref{eq_swap} holds, we say $\left(P_{\bbs}(x,y)\right)_{j}$ and $\left(P_{\bbs}^{*}(x,y)\right)_{j}$ are \textit{swappable}. 
From symmetry, we obtain that $\left(P_{\bbs}(x,y)\right)_{j}$ and $\left(P_{\bbs}^{*}(x,y)\right)_{j}$ are swappable if and only if $\left(P_{\bbs}(x,y)\right)_{n-j}$ and $\left(P_{\bbs}^{*}(x,y)\right)_{n-j}$ are swappable.

Lastly, to ensure that $Q_1(x,y)=Q_2^{*}(x,y)$, we need to swap $\left(P_{\bbs}(x,y)\right)_{j}$ and $\left(P_{\bbs}^{*}(x,y)\right)_{j}$, $\left(P_{\bbs}(x,y)\right)_{n-j}$ and $\left(P_{\bbs}^{*}(x,y)\right)_{n-j}$ simultaneously.

Putting everything together, we get the following lemma.
\begin{lemma}\label{lem_notunique}
  Let $\bbs$ be a string and let $P_{\bbs}(x,y)$ be its generating polynomial. 
  Then there exists a string $\bbt$ such that $\bbt\ne \bbs,\bbss$ and $\mathcal{M}(\bbt)=\mathcal{M}(\bbs)$ if there exists some $1\le j\le\lfloor\frac{n}{2}\rfloor$ such that the following three conditions hold:
  \renewcommand{\labelenumi}{(\roman{enumi})}
  \begin{enumerate}
    \item \Cref{eq_swap} holds.
    \item $a_j\ne w-a_{n-j}$, i.e, $\wth\left(\bbs_{1}^{j}\right)\ne \wth\left(\bbs_{n-j+1}^{n}\right)$.
    \item There exists some $k\ne j,n-j$ such that $\left(P_{\bbs}(x,y)\right)_{k}\ne \left(P_{\bbs}^{*}(x,y)\right)_{k}$, i.e., $\wth\left(\bbs_{1}^{k}\right)\ne \wth\left(\bbs_{n-k+1}^{n}\right)$.
  \end{enumerate}
\end{lemma}
\begin{IEEEproof}
  By Lemma \ref{lem_confusion}, there exists a string $\bbt$ such that $\mathcal{M}(\bbt)=\mathcal{M}(\bbs)$ if and only if there is a polynomial $Q(x,y)$ with coefficients in $\{0,1\}$ such that $P_{\bbs}(x,y)+P_{\bbs}^{*}(x,y)=Q(x,y)+Q^{*}(x,y)$ and $Q(x,y)$ indeed gives a string. Condition (i) ensures the existence of such $Q(x,y)$.
  The last two conditions exclude the cases $\bbt=\bbs$ and $\bbt=\bbss$. 
  Indeed, let $Q(x,y)$ be obtained by swapping $\left(P_{\bbs}(x,y)\right)_{j}$ and $\left(P_{\bbs}^{*}(x,y)\right)_{j}$, $\left(P_{\bbs}(x,y)\right)_{n-j}$ and $\left(P_{\bbs}^{*}(x,y)\right)_{n-j}$ simultaneously. Let $\bbt$ be the string corresponding to $Q(x,y)$. Then $\bbt\ne \bbs$ is equivalent to $\left(P_{\bbs}(x,y)\right)_{j}\ne\left(P_{\bbs}^{*}(x,y)\right)_{j}$ and $\left(P_{\bbs}(x,y)\right)_{n-j}\ne\left(P_{\bbs}^{*}(x,y)\right)_{n-j}$, which is equivalent to condition (\Rnum{2}). Lastly, $\bbt\ne \bbss$ is equivalent to condition (\Rnum{3}).
\end{IEEEproof}

With Lemma \ref{lem_notunique}, we can now describe the explicit structure of a class of strings that cannot be uniquely reconstructed up to reversal.
\begin{theorem}\label{thm_notunique}
  Let $\bbs=s_1\cdots s_n\in\Sigma_2^n$, where $n\ge 2$. 
  Then $\bbs$ cannot be uniquely reconstructed up to reversal if it has one of the following forms:
  \renewcommand{\labelenumi}{(\roman{enumi})}
  \begin{enumerate}
    \item $\bbs=s_1\cdots s_{j-1}10s_{j+2}\cdots s_{n-j-1}10s_{n-j+2}\cdots s_n$ and $\wth\left(\bbs_1^{j}\right)=\wth\left(\bbs_{n-j+1}^{n}\right)+1$ for some $1\le j\le \frac{n}{2}$ when $n$ is even, or $1\le j\le \lfloor\frac{n}{2}\rfloor-1$ when $n$ is odd.
    \item $\bbs=s_1\cdots s_{j-1}01s_{j+2}\cdots s_{n-j-1}01s_{n-j+2}\cdots s_n$ and $\wth\left(\bbs_1^{j}\right)=\wth\left(\bbs_{n-j+1}^{n}\right)-1$ for some $1\le j\le \frac{n}{2}$ when $n$ is even, or $1\le j\le \lfloor\frac{n}{2}\rfloor-1$ when $n$ is odd.
  \end{enumerate}
  And there exists some $k\ne j,n-j$ such that $\wth\left(\bbs_{1}^{k}\right)\ne \wth\left(\bbs_{n-k+1}^{n}\right)$.
\end{theorem}
\begin{IEEEproof}
 Let $P_{\bbs}(x,y)$ be a generating polynomial for a string $\bbs$.
 Recall that $\bbs$ cannot be uniquely reconstructed up to reversal if the three conditions in Lemma \ref{lem_notunique} hold. 
 In order to prove the theorem, we show that the two conditions in this theorem are equivalent to the three conditions in \Lref{lem_notunique}. It is easy to verify that if $\bbs$ satisfies (\Rnum{1}) or (\Rnum{2}) in this theorem, it also satisfies the three conditions in Lemma \ref{lem_notunique}.
 
 Next, we will show that we can obtain
\Cref{thm_notunique} from \Lref{lem_notunique}. First, notice that since $P_{\bbs}(x,y)$ is a generating polynomial,  $a_{n-j-1}\le a_{n-j+1}\le a_{n-j-1}+2$. Now consider the following possible cases.

\begin{description}
 \item[Case (1):] $a_{n-j+1}=a_{n-j-1}=a_{n-j}$. Then the first row of \Cref{eq_swap} implies that $a_j=w-a_{n-j}$, which contradicts condition (\Rnum{2}) in Lemma \ref{lem_notunique}.

 \item[Case (2):] $a_{n-j+1}=a_{n-j-1}+2$. Then $a_{n-j}=a_{n-j-1}+1$. The second row of \Cref{eq_swap} implies that $a_j=w-a_{n-j+1}+1=w-a_{n-j-1}-1=w-a_{n-j}$, which contradicts condition (\Rnum{2}) in Lemma \ref{lem_notunique}.

 \item[Case (3):] $a_{n-j+1}=a_{n-j-1}+1$. Then the first row of \Cref{eq_swap} implies that $a_j=w-a_{n-j-1}$ or $a_{j}=w-a_{n-j-1}-1$. 
 \begin{itemize}
   \item For the case $a_j=w-a_{n-j-1}$: if $a_{n-j}=a_{n-j-1}$, then $a_j=w-a_{n-j}$, which contradicts condition (\Rnum{2}) in Lemma \ref{lem_notunique}. So $a_{n-j}=a_{n-j-1}+1=w-a_{j}+1$, which is equivalent to $\wth\left(\bbs_1^{j}\right)=\wth\left(\bbs_{n-j+1}^{n}\right)+1$. From the third and fourth rows of \Cref{eq_swap}, we get $a_{j+1}=a_{j}=a_{j-1}+1$. Combing condition (\Rnum{3}) in Lemma \ref{lem_notunique}, we have proved (\Rnum{1}).
   \item $a_j=w-a_{n-j-1}-1$. If $a_{n-j}=a_{n-j+1}$, then $a_j=w-a_{n-j}$, which contradicts condition (\Rnum{2}) in Lemma \ref{lem_notunique}. So $a_{n-j}=a_{n-j-1}=w-a_{j}-1$, which is equivalent to $\wth\left(\bbs_1^{j}\right)=\wth\left(\bbs_{n-j+1}^{n}\right)-1$.. From the third and fourth rows of \Cref{eq_swap}, we get $a_{j-1}=a_{j}=a_{j+1}-1$. Combing condition (\Rnum{3}) in Lemma \ref{lem_notunique}, we have proved (\Rnum{2}).
 \end{itemize}
 \end{description}
 Note that if $n$ is odd and $j=\lfloor\frac{n}{2}\rfloor$, then $n-j=j+1$. So neither of the above two cases can hold. 
\end{IEEEproof}

\begin{remark}\label{rmk_notunique}
The following observations will be helpful for the analysis in this paper.
  \begin{itemize}
    \item According Remark \ref{rmk_weights}, to find the desired $k$ in \Cref{thm_notunique}, we only need to consider $1\le k\le\left\lfloor\frac{n}{2}\right\rfloor$.
    \item Note that a string $\bbs$ satisfies (\Rnum{1}) of \Cref{thm_notunique} if and only if $\bbss$ satisfies (\Rnum{2}) of \Cref{thm_notunique}.
    \item For any string $\bbs\in\Sigma_2^n$, we have $\wth\left(\bbs_1^{j}\right)$$=$ $\wth\left(\bbs_{n-j+1}^{n}\right)+1$ if and only if $\wth\left(\bbs_1^{n-j}\right)$$=$ $\wth\left(\bbs_{j+1}^{n}\right)+1$.
    \item If $\bbs$ satisfies (\Rnum{1}) or (\Rnum{2}) of \Cref{thm_notunique}, then $\wth\left(\bbs_1^{j-1}\right)= \wth\left(\bbs_{n-j+2}^{n}\right)$.
  \end{itemize}
\end{remark}

We present two examples to illustrate \Cref{thm_notunique}.
\begin{example}\label{exmp_notunique}
\begin{enumerate}
  \item Consider $n=6$, which implies $1\le j \le 3$. If $j=1$, using \Cref{thm_notunique} we can only consider $k=3$, since $\wth(\bbs_1^2)=\wth(\bbs_5^6)$. 
    Then from (\Rnum{1}) in \Cref{thm_notunique} we get two strings: $101010$ and $100110$, and from (\Rnum{2}) in \Cref{thm_notunique} we get two strings: $010101$ and $011001$. It can be directly verified that $\mathcal{M}(101010)=\mathcal{M}(100110)=\mathcal{M}(010101)=\mathcal{M}(011001)$. 
    
    If $j=2$, we cannot find a desired $k\neq j,n-j$ for which $\wth(\bbs_1^k)=\wth(\bbs_{6-k+1}^6)$. 
    
    If $j=3$, then we can only consider $k=1$. Then from (\Rnum{1}) in \Cref{thm_notunique} we get two strings: $101010$ and $011001$, and from (\Rnum{2}) in \Cref{thm_notunique} we get two strings: $010101$ and $100110$. 
    Thus, for $n=6$, there are $4$ strings that cannot be uniquely reconstructed up to reversal and they share the same prefix-suffix composition multiset.

   \item Consider $n=7$, which implies $1\le j \le 2$. If $j=1$, using \Cref{thm_notunique} we can only consider $k=3$. From (\Rnum{1}) in \Cref{thm_notunique} we get four strings: $1011010, 1010010, 1001110,$ and $1000110$. The strings we get from (\Rnum{2}) in \Cref{thm_notunique} are nothing but the reversals of those we get from (\Rnum{1}): $0101101, 0100101, 0111001$ and $0110001$. It can be directly verified that 
   \[\mathcal{M}(1011010)=\mathcal{M}(0101101)= \mathcal{M}(1001110)=\mathcal{M}(0111001),\] 
   and that 
   \[\mathcal{M}(1010010)=\mathcal{M}(0100101)=\mathcal{M}(1000110)=\mathcal{M}(0110001).\]
   
   If $j=2$, we cannot find a desired $k$. 
   Thus, for $n=7$, there are $8$ strings that cannot be uniquely reconstructed up to reversal and among them, four share a common prefix-suffix composition multiset and the other four share another common prefix-suffix composition multiset.
\end{enumerate}
\end{example}

The above two examples imply that when $n=6,7$, there exist length-$n$ strings that cannot be uniquely reconstructed up to reversal. The following corollary asserts that this is true for all $n\ge 6$. 

\begin{corollary}
  For any $n\ge 6$, there always exists a string of length $n$ that cannot be uniquely reconstructed up to reversal.
\end{corollary}

\begin{IEEEproof}
  Let $\bbs=100s_4\cdots s_{n-3}110$. Then $\bbs$ satisfies (\Rnum{1}) in \Cref{thm_notunique} with $j=1$ and $k=3$.
\end{IEEEproof}

\subsection{The General Case}
\Cref{lem_notunique} only gives a sufficient condition for a string to be uniquely reconstructible up to reversal. The following example gives a string that can be uniquely reconstructed up to reversal, but does not satisfy \Cref{lem_notunique}.
\begin{example}\label{examp_fl}
Let $\bbs=110100100110\in\Sigma_2^{12}$. Then $\bbss=011001001011$ and
$$
\begin{array}{l}
   P_{\bbs}\parenv{x,y}= 1+x+x^2+x^2y+x^3y+x^3y^2+x^3y^3+x^4y^3+x^4y^4+x^4y^5+x^5y^5+x^6y^5+x^6y^6,  \\
   P_{\bbs}^{*}\parenv{x,y}=1+y+xy+x^2y+x^2y^2+x^2y^3+x^3y^3+x^3y^4+x^3y^5+x^4y^5+x^4y^6+x^5y^6+x^6y^6,\\
   D=\mathset{1,2,4,5,7,8,10,11}.
\end{array}
$$
It is easy to see that $\bbs$ does not satisfy the conditions in \Cref{thm_notunique}. Let $S=\mathset{1,2}$, then
$$
\begin{array}{l}
   Q_{S,1}\parenv{x,y}= 1+y+xy+x^2y+x^3y+x^3y^2+x^3y^3+x^4y^3+x^4y^4+x^4y^5+x^4y^6+x^5y^6+x^6y^6,  \\
   Q_{S,2}\parenv{x,y}=1+x+x^2+x^2y+x^2y^2+x^2y^3+x^3y^3+x^3y^4+x^3y^5+x^4y^5+x^5y^5+x^6y^5+x^6y^6.
\end{array}
$$
Let $\bbt$ be the string corresponding to $Q_{S,1}\parenv{x,y}$. Then $\bbt=011100100011$ and $\cM(\bbt)=\cM(\bbs)$. Clearly, $\bbt\ne\bbs$ and $\bbt\ne\bbss$.
\end{example}

In the above example, if we let $S=\mathset{1}$, then
$$
\begin{array}{l}
 Q_{S,1}\parenv{x,y}= 1+y+x^2+x^2y+x^3y+x^3y^2+x^3y^3+x^4y^3+x^4y^4+x^4y^5+x^5y^5+x^5y^6+x^6y^6,  \\
   Q_{S,2}\parenv{x,y}=1+x+xy+x^2y+x^2y^2+x^2y^3+x^3y^3+x^3y^4+x^3y^5+x^4y^5+x^4y^6+x^6y^5+x^6y^6.
\end{array}
$$
However, neither $Q_{S,1}\parenv{x,y}$ nor $Q_{S,2}\parenv{x,y}$ is a generating polynomial. Therefore, we should first find the conditions under which $Q_{S,1}\parenv{x,y}$ is a generating polynomial.

Let $I\subseteq\mathset{1,\ldots,n}$. If there are $1\le i\le j\le n$ such that $I=\mathset{i,i+1,\ldots,j}$, we say $I$ is an \emph{interval}. Any nonempty subset $S$ of $D\cap\mathset{1,\ldots,\floorenv{\frac{n}{2}}}$ is a union of maximal intervals, i.e., $S=\bigcup_{i}I_i$, where each $I_i$ is an interval and $I_i\cup I_j$ is not an interval whenever $i\ne j$. For example, in Example \ref{examp_fl}, let $S_1=\{1,2\}$ and $S_2=\{1,2,4,5\}$. Then $S_1$ is a maximal interval and $S_2=I_1\cup I_2$, where $I_1=\{1,2\}$ and $I_2=\{4,5\}$. Both of $I_1$ and $I_2$ are maximal intervals since $I_1\cup I_2$ is not an interval.
\begin{lemma}\label{lem_interval}
Let $\bbs$ be a string of length $n$ with generating polynomial $P_{\bbs}(x,y)$.
Let $S=\bigcup_{i}I_i$ be a nonempty subset of $D\cap\mathset{1,\ldots,\floorenv{\frac{n}{2}}}$ and suppose that $Q_{S,1}\parenv{x,y}$ gives a string which is different from $\bbs$ and $\bbss$. Then there exist some $i$ such that $Q_{I_i,1}\parenv{x,y}$ also gives a string which is different from $\bbs$ and $\bbss$.
\end{lemma}
\begin{IEEEproof}
   Notice that if $Q_{S,1}\parenv{x,y}$ gives a string, then this string is different from $\bbs$ and $\bbss$ if and only if there exists a $j\in D\cap\mathset{1,\ldots,\floorenv{\frac{n}{2}}}$ such that $j\notin S$. Now the conclusion follows.
\end{IEEEproof}

By the above lemma, we can assume in our analysis that $S$ is an internal. \Lref{lem_legal} below is a generalization of \Lref{lem_notunique} (\Rnum{1}).
\begin{lemma}\label{lem_legal}
Let $\bbs$ be a string and let $P_{\bbs}(x,y)$ be its generating polynomial.
Let $S=\mathset{i_1,\ldots,i_2}$ is an interval of $D\cap\mathset{1,\ldots,\floorenv{\frac{n}{2}}}$. Then $Q_{S,1}(x,y)$ gives a string if and only if $\parenv{P_{\bbs}(x,y)}_{i_1-1}=\parenv{P_{\bbs}^{*}(x,y)}_{i_1-1}$ and $\parenv{P_{\bbs}(x,y)}_{i_2+1}=\parenv{P_{\bbs}^{*}(x,y)}_{i_2+1}$.
\end{lemma}
\begin{IEEEproof}
The $\Leftarrow$ direction is clear. So we only need to prove the $\Rightarrow$ direction.

According to the definition of $D$, we have $\parenv{P_{\bbs}(x,y)}_{j}\ne\parenv{P_{\bbs}^{*}(x,y)}_{j}$, i.e., $\wth\parenv{\bbs_1^j}\ne\wth\parenv{(\bbss)_1^j}$, for all $j\in S$. Suppose $\parenv{P_{\bbs}(x,y)}_{i_1-1}=x^ay^b$, $\parenv{P_{\bbs}^{*}(x,y)}_{i_1-1}=x^cy^d$, $\parenv{P_{\bbs}(x,y)}_{i_1}=x^{a^\prime}y^{b^\prime}$ and $\parenv{P_{\bbs}^{*}(x,y)}_{i_1}=x^{c^\prime}y^{d^\prime}$. Since $Q_{S,1}(x,y)$ gives a string, we have
\begin{equation}\label{eq_four}
\left\{
\begin{array}{l}
   a\le a^\prime\le a+1,   \\
     c\le a^\prime\le c+1,\\
     a\le c^\prime\le a+1,\\
     c\le c^\prime\le c+1.
\end{array}
\right.
\end{equation}
Notice that $a^\prime>c^\prime$ or $a^\prime<c^\prime$ since $\wth\parenv{\bbs_1^{i_1}}\ne\wth\parenv{(\bbss)_1^{i_1}}$. Then from \Cref{eq_four} we can conclude that $a=c$ and hence $\parenv{P_{\bbs}(x,y)}_{i_1-1}=\parenv{P_{\bbs}^{*}(x,y)}_{i_1-1}$.

In a similar way we can also prove $\parenv{P_{\bbs}(x,y)}_{i_2+1}=\parenv{P_{\bbs}^{*}(x,y)}_{i_2+1}$.
\end{IEEEproof}

The following theorem is a generalization of \Lref{lem_notunique}.
\begin{theorem}\label{thm_notuniquegeneral}
Let $\bbs$ be a string and let $P_{\bbs}(x,y)$ be its generating polynomial given in \Cref{eq_generating}. 
  Then there exists a string $\bbt$ such that $\bbt\ne \bbs,\bbss$ and $\mathcal{M}(\bbt)=\mathcal{M}(\bbs)$ if and only if there exists an interval $S=\mathset{i_1,\ldots,i_2}$ of $D\cap\mathset{1,\ldots,\floorenv{\frac{n}{2}}}$ such that the following two conditions hold:
  \renewcommand{\labelenumi}{(\roman{enumi})}
  \begin{enumerate}
    \item $\parenv{P_{\bbs}(x,y)}_{i_1-1}=\parenv{P_{\bbs}^{*}(x,y)}_{i_1-1}$ and $\parenv{P_{\bbs}(x,y)}_{i_2+1}=\parenv{P_{\bbs}^{*}(x,y)}_{i_2+1}$.
    \item There exists some $k\in D\cap\mathset{1,\ldots,\floorenv{\frac{n}{2}}}\setminus S$ such that $\left(P_{\bbs}(x,y)\right)_{k}\ne \left(P_{\bbs}^{*}(x,y)\right)_{k}$, i.e., $\wth\left(\bbs_{1}^{k}\right)\ne \wth\left(\bbs_{n-k+1}^{n}\right)$.
  \end{enumerate}
\end{theorem}
\begin{IEEEproof}
We first prove the $\Rightarrow$ direction. By \Lref{lem_interval}, we can choose a $\bbt$ whose generating polynomial corresponds in interval.  Then (\Rnum{1}) follows from \Lref{lem_legal} and (\Rnum{2}) follows the fact that $\bbt\ne\bbss$.

Now we prove the $\Leftarrow$ direction.  Condition (\Rnum{1}) and \Lref{lem_legal} ensure that $Q_{S,1}(x,y)$ does give a string. Let $\bbt$ be the string with generating polynomial $Q_{S,1}(x,y)$. The definition of $D$ ensures $\bbt\ne\bbs$. Condition (\Rnum{2}) ensures $\bbt\ne\bbss$.
\end{IEEEproof}

With \Cref{thm_notunique} and \Cref{thm_notuniquegeneral}, we can obtain that for $n\le 5$, every string of length $n$ is uniquely reconstructible up to reversal. 
\begin{corollary}\label{cor_unique}
\begin{itemize}
  \item When $n=6$ or $7$, all the string that are not uniquely reconstructible are listed in Example \ref{exmp_notunique}.
  \item  When $n\le 5$, each string of length $n$ can be uniquely reconstructed up to reversal.
\end{itemize}
\end{corollary}
\begin{IEEEproof}
First, we notice that $\abs{S}\le 2$ since $\floorenv{\frac{n}{2}}\le 2$. If $S=\{1,2\}$, we can not find the desired $k$ in \Cref{thm_notuniquegeneral} (\Rnum{2}). Therefore $\abs{S}=1$ and it suffices to apply \Cref{thm_notunique}. So the first conclusion is proved. Next we prove the second conclusion.
  \begin{enumerate}
    \item[] $n=1$: This case is trivial since there are only two strings $1$ and $0$.
    \item[] $n=2$: In \Cref{thm_notunique}, we can only take $j=1$ and the candidate strings are $10$ and $01$. In either case, we cannot find a desired $k$.
    \item[] $n=3$: The requirement for $j$ is $1\le j\le\lfloor\frac{3}{2}\rfloor-1=0$, which is impossible.
    \item[] $n=4$: We can only take $j=1$ or $j=2$. If $j=1$, the candidate strings are $1010$ and $0101$. In either case, we cannot find a desired $k\ne j,n-j$ such that $\wth\left(\bbs_{1}^{k}\right)\ne \wth\left(\bbs_{n-k+1}^{n}\right)$. If $j=2$, the candidate strings are $0100, 0101, 1100,1101$ and their reversals. If $\bbs=0100,0101,1101$ or their reversals, we cannot find $k\ne j,n-j$ such that $\wth\left(\bbs_{1}^{k}\right)\ne \wth\left(\bbs_{n-k+1}^{n}\right)$. If $\bbs=1100$, then $\wth\left(\bbs_{1}^{2}\right)=\wth\left(\bbs_{3}^{4}\right)+2$, which is not of the form (\Rnum{1}) nor (\Rnum{2}) in \Cref{thm_notunique}. Similarly, if $\bbs=0011$, then $\wth\left(\bbs_{1}^{2}\right)=\wth\left(\bbs_{3}^{4}\right)-2$, which is not of the form (\Rnum{1}) nor (\Rnum{2}) in \Cref{thm_notunique}..
    \item[] $n=5$: We can only take $j=1$. The candidate strings are $10010,10110$ and their reversals. In any case, we cannot find $k\ne j,n-j$ such that $\wth\left(\bbs_{1}^{k}\right)\ne \wth\left(\bbs_{n-k+1}^{n}\right)$.
  \end{enumerate}
\end{IEEEproof}

\section{Strings that can be uniquely reconstructed up to reversal}\label{sec_unique}
In the previous section we obtained that for any $n\ge 6$, there always exists a string of length $n$ that cannot be uniquely reconstructed up to reversal. 
Two natural questions arise:
\begin{enumerate}
  \item For a given $n$, how many strings of length $n$ are there which can be uniquely reconstructed up to reversal?
  \item Is there a unified method to construct all the uniquely reconstructible strings?
\end{enumerate}
This section is devoted to solving these questions. Let $\mathcal{U}(n)\subseteq\Sigma_2^n$ be the set of all $n$-length strings that can be uniquely reconstructed up to reversal.

For any $\bbs\in\Sigma_2^n$, we can always find $0=j_0<j_1<\cdots <j_{\ell}\le \lfloor\frac{n}{2}\rfloor$, where $\ell\ge 0$, such that $\left(P_{\bbs}(x,y)\right)_{j}=\left(P_{\bbs}^{*}(x,y)\right)_{j}$ for all $j=j_i,n-j_{i}$, where $0\le i\le\ell$, and $\left(P_{\bbs}(x,y)\right)_{j}\ne\left(P_{\bbs}^{*}(x,y)\right)_{j}$ for all $j\ne j_i,n-j_{i}$. 
In other words, $\wth\left(\bbs_1^j\right)=\wth\left(\bbs_{n-j+1}^n\right)$ for all $j=j_i,n-j_{i}$, and $\wth\left(\bbs_1^j\right)\ne \wth\left(\bbs_{n-j+1}^n\right)$ for all $j\ne j_i,n-j_{i}$, which is the same as $\wth\left(\bbs_1^j\right)=\wth\left((\bbss)_{1}^{j}\right)$ for all $j=j_i,n-j_{i}$, and $\wth\left(\bbs_1^j\right)\ne \wth\left((\bbss)_{1}^{j}\right)$ for all $j\ne j_i,n-j_{i}$ (see Remark \ref{rmk_weights}). Clearly, $\mathset{j_1,\ldots,j_{\ell}}=\mathset{1,\ldots,\floorenv{\frac{n}{2}}}\setminus D$, where $D$ is defined at the beginning of \Cref{sec_notunique}.

We give some examples to illustrate the above idea.
\begin{example}\label{examp_partition}
  \begin{enumerate}
    \item Let $n=6$ and $\bbs=101010$. Then $P_{\bbs}(x,y)=1+x+xy+x^2y+x^2y^2+x^3y^2+x^3y^3$ and $P_{\bbs}^{*}(x,y)=1+y+xy+xy^2+x^2y^2+x^2y^3+x^3y^3$. We have  $\ell=1$ and $j_1=2$.
    \item Let $n=6$ and $\bbs=010111$. Then $P_{\bbs}(x,y)=1+y+xy+xy^2+x^2y^2+x^3y^2+x^4y^2$ and $P_{\bbs}^{*}(x,y)=1+x+x^2+x^3+x^3y+x^4y+x^4y^2$ which means that $\ell=0$.
    \item Let $n=7$ and $\bbs=1011010$. Then $P_{\bbs}(x,y)=1+x+xy+x^2y+x^3y+x^3y^2+x^4y^2+x^4y^3$ and $P_{\bbs}^{*}(x,y)=1+y+xy+xy^2+x^2y^2+x^3y^2+x^3y^3+x^4y^3$. We have $\ell=1$ and $j_1=2$.
  \end{enumerate}
\end{example}

Next we will solve the two questions raised at the beginning of this section. We will not give the general construction of $\cU(n)$ immediately. Instead, we first consider two special cases, which not only contain interesting results themselves, but also provide us a better understanding of the ideas behind.
\subsection{The case $\ell=0$}
We consider the case $\ell=0$. That is to say, $\wth\left(\bbs_1^j\right)\ne \wth\left(\bbs_{n-j+1}^n\right)$ for all $1\le j\le n-1$. The authors of \cite{wz3} constructed the set $\mathcal{S}_R(n)$ as follows:
\begin{equation}\label{eq_srn}
  \begin{array}{rl}
    \mathcal{S}_R(n)&=\left\{\bbs\in\Sigma_2^n:\;s_1=0,s_n=1,\text{ and }\exists I\subset\{1,\ldots,n\}\text{ such that}\right.\\
    &\quad\;\;s_i\ne s_{n-i+1}\text{ for all }i\in I; s_i=s_{n-i+1}\text{ for all }i\notin I;\\
    &\quad\;\;\left.\bbs_{\mathset{1,\dots,\frac{n}{2}}\cap I}\text{ is a Catalan-Bertrand string}\right\}
  \end{array}
\end{equation}
when $n$ is even, and
\begin{align*}
    \cS_R(n)&=\bigcup_{\bbs\in\cS_R(n-1)}\mathset{\bbs_{1}^{\frac{n-1}{2}}\,0\,\bbs_{\frac{n+1}{2}}^{n-1}\;\; , \;\; \bbs_{1}^{\frac{n-1}{2}}\,1\,\bbs_{\frac{n+1}{2}}^{n-1}}
\end{align*}
when $n$ is odd. Here, a Catalan-Bertrand string is a binary string such that every prefix contains strictly more $0$s than $1$s.

\begin{lemma}\label{lem_catlan}
Let $\bbs\in\Sigma_2^n$. Then $\wth\left(\bbs_1^j\right)\ne \wth\left(\bbs_{n-j+1}^n\right)$ for all $1\le j\le n-1$ if and only if $\bbs$ or $\bbss$ is in $\mathcal{S}_R(n)$.
\end{lemma}
\begin{IEEEproof}
  The $\Leftarrow$ direction follows from \cite[Lemma 1]{wz3}. For the $\Rightarrow$ direction, without loss of generality assume $s_1=0$ and $s_n=1$. 
  Let $I\subseteq\{1,\dots,n\}$ such that $s_i\ne s_{n-i+1}$ if and only if $i\in I$. Notice that $1,n\in I$. 
  Since any $i\notin I$ does not affect $\parenv{\wth (\bbs_1^j)-\wth(\bbs_{n-j+1}^n)}$, together with the assumptions that  $s_1=0,s_n=1$ and that  $\wth\left(\bbs_1^j\right)\ne \wth\left(\bbs_{n-j+1}^n\right)$ for all $1\le j\le n-1$, we conclude that $\wth\left(\bbs_1^j\right)< \wth\left(\bbs_{n-j+1}^n\right)$ for all $1\le j\le n-1$. It follows that $\bbs_{\left[\frac{n}{2}\right]\cap I}$ is a Catalan-Bertrand string.
\end{IEEEproof}

The following corollary is a direct result from Lemma \ref{lem_catlan}.
\begin{corollary}
  For a given $n$, let $f(n)$ be the number of strings $\bbs$ of length $n$ such that $\wth\left(\bbs_1^j\right)\ne \wth\left(\bbs_{n-j+1}^n\right)$ for all $1\le j\le n-1$. Then
  \begin{equation}\label{eq_f}
    f(n)=
    \begin{cases}
      2\left|\mathcal{S}_R(n)\right|, & \mbox{if } n\ge 2, \\
      2, & \mbox{if } n=1.
    \end{cases}
  \end{equation}
\end{corollary}
The value of $\left|\mathcal{S}_R(n)\right|$ was given in \cite{wz3}:
\begin{equation}\label{eq_sizeofS}
  \left|\mathcal{S}_R(n)\right|=
  \begin{cases}
   \mathop{\sum}\limits_{i=0}^{\frac{n}{2}-1}\binom{\frac{n}{2}-1}{i}2^{\frac{n}{2}-1-i}\binom{i}{\lfloor\frac{i}{2}\rfloor} \ge \frac{2^{n-3}}{\sqrt{\pi n}} , & \mbox{if } n\text{ is even}, \\
    2\left|\mathcal{S}_R(n-1)\right|, & \mbox{if } n\text{ is odd}.
  \end{cases}
\end{equation}
Here we define $\mathcal{S}_R(0)\triangleq\left\{\emptyset\right\}$ and hence $\left|\mathcal{S}_R(0)\right|= 1$.
Therefore, 
$$
\mathop{\lim}\limits_{n\rightarrow\infty}\frac{\log_2\left|\mathcal{S}_R(n)\cup\mathcal{S}_R(n)^{*}\right|}{n}=1,
$$
where $\mathcal{S}_R(n)^{*}\triangleq\left\{\bbs:\;\bbss\in\mathcal{S}_R(n)\right\}$.

\begin{lemma}\label{lem_notequal}
  Let $\bbs\in\Sigma_2^n$ be a string such that $\wth\left(\bbs_1^j\right)\ne \wth\left(\bbs_{n-j+1}^n\right)$ for all $1\le j\le n-1$. Then $\bbs$ is uniquely reconstructible up to reversal.
\end{lemma}
\begin{IEEEproof}
  When $n=1$, $\bbs=0$ or $1$. So the conclusion is trivial. When $n\ge2$, we apply \Cref{thm_notuniquegeneral}.
  
  If $n=2$, $\bbs=01$ or $10$. we can only choose $S=\{1\}$, which does not satisfy condition (\Rnum{1}) of \Cref{thm_notuniquegeneral}.
  When $n\ge 3$, we have $\floorenv{\frac{n}{2}}+1<n$. Therefore, it is impossible to find an interval $S$ satisfying condition (\Rnum{1}) of \Cref{thm_notuniquegeneral}.
\end{IEEEproof}

There is another way to show the correctness of \Lref{lem_notequal}. Let $\bbs$ be given in \Lref{lem_notequal}. We can conclude that $\wth\left(\bbs_1^j\right)> \wth\left(\bbs_{n-j+1}^n\right)$ for all $1\le j\le n-1$, or $\wth\left(\bbs_1^j\right)< \wth\left(\bbs_{n-j+1}^n\right)$ for all $1\le j\le n-1$. So when given $s_1$, we can know the compositions of all prefixes of $\bbs$ and thus reconstruct $\bbs$. There are two cases for $s_1$, which means that we can get two strings and they are $\bbs$ and $\bbss$.

To conclude the case of $\ell=0$, we give the following corollary, which is an immediate result of \Lref{lem_catlan} and \Lref{lem_notequal}, presents a set of strings which can be uniquely reconstructed up to reversal.
\begin{corollary}
    For all $n\ge 1$, any string in $\mathcal{S}_R(n)\cup\mathcal{S}_R(n)^{*}$ can be uniquely reconstructed up to reversal.
\end{corollary}

\subsection{General $\ell$}
Now we are ready to deal with general $\ell$. 
For convenience, in the sequel we denote $j_{\ell+1}:=n-j_{\ell}$ (recall that $j_{\ell}\leq \floorenv{n/2}$).
The following theorem is a reformulation of \Cref{thm_notuniquegeneral}. 
\begin{theorem}\label{thm_notunique2}
  Let $\bbs\in\Sigma_2^n$ and let $j_0,\ldots,j_{\ell},j_{\ell+1}$ be defined as above. Then $\bbs$ cannot be uniquely reconstructed up to reversal if and only if there exist two different integers $i,i^{\prime}\in\{1,\ldots,\ell+1\}$ such that $j_i-j_{i-1}\ge2$ and $j_{i^{\prime}}-j_{i^{\prime}-1}\ge2$. In particular, we have $\ell\ge 1$.
\end{theorem}
Indeed, in \Cref{thm_notuniquegeneral}, we can choose $S=\{j_{i-1}+1,\ldots,j_i-1\}$. Then by the definition of $j_0,j_1,\ldots,j_{\ell+1}$, condition (\Rnum{1}) of \Cref{thm_notuniquegeneral} is satisfied. Since $j_{i^{\prime}}-j_{i^{\prime}-1}\ge2$, we can choose a $k$ with $j_{i^{\prime}-1}<k<j_{i^{\prime}}$. Besides, we have $k\notin S$ because $i\ne i^{\prime}$. Then condition (\Rnum{2}) of \Cref{thm_notuniquegeneral} is satisfied.

The following is a construction of a subset of $\mathcal{U}(n)$ for any given $n\ge 1$: 
\begin{equation*}
\begin{array}{rl}
  \mathcal{U}^{\prime}(n)&=\left\{\bm{u}_1\cdots\bm{u}_{\ell}\bm{u}_{\ell+1}\bm{u}_{\ell}\cdots\bm{u}_{1}:\;\ell\ge 0,j_0=0;\right.\\
  &\quad\;\;1\le j_1<\cdots j_{\ell}\le\lfloor\frac{n}{2}\rfloor, j_{\ell+1}=n-j_{\ell};\\
  &\quad\;\;\bm{u}_i\text{ or }\bm{u}_i^{*}\in\mathcal{S}_R(j_i-j_{i-1})\text{ for all }1\le i\le\ell+1;\\
  &\quad\;\;\left.j_i-j_{i-1}\ge 2\text{ for at most one }i
  .\right\}
\end{array}
\end{equation*}

\begin{theorem}
  For any $n\ge 1$, we have $\mathcal{U}^{\prime}(n)\subseteq\mathcal{U}(n)$.
\end{theorem}
\begin{IEEEproof}
  The conclusion follows from \Cref{thm_notunique2} and Lemma \ref{lem_catlan}. Indeed, Lemma \ref{lem_catlan} ensures that $\wth(\bbs_1^j)\neq \wth(\bbs_{n-j+1}^n)$ except for the selected $j_i$s.
\end{IEEEproof}
\begin{remark}\label{rmk_unique}
Take $\ell=0$, we can see that $\cS_R(n)\cup\cS_R(n)^*\subseteq \cU^{\prime}(n)$.
Notice that for $n\ge 6$, $\mathcal{S}_R(n)\cup\mathcal{S}_R(n)^{*}$ is not necessarily a equal to  $\mathcal{U}^{\prime}(n)$.
\end{remark}

It is possible to provide a formula for $\abs{\cU^{\prime}(n)}$. 
For $n\ge1$, define
\begin{align} 
\label{eq:P_def}
  \mathcal{P}(n)&\triangleq\left\{\bbj=(j_0,j_1,\ldots,j_{\ell+1}):\ell\ge 0;j_0=0; 1\le j_1<\cdots<j_{\ell}\le\left\lfloor\frac{n}{2}\right\rfloor,\right.\\ \nonumber
  &\quad\;\;\left.j_{\ell+1}=n-j_{\ell};\text{there is at most one }i\text{ such that }j_i-j_{i-1}\ge2\right\}.
\end{align}
The size of $\mathcal{U}^{\prime}(n)$ is
\[
\abs{\cU^{\prime}(n)}=\sum_{\ell=0}^{\floorenv{\frac{n}{2}}}\sum_{\bbj\in\cP(n)} \prod_{i=1}^{\ell+1}f(j_i-j_{i-1}),\]
where $f(m)$ is given in \Cref{eq_f} for $m\ge 1$ and $f(0)\triangleq 1$. 
Since $\mathcal{S}_R(n)\cup\mathcal{S}_R(n)^{*}\subseteq\mathcal{U}^{\prime}(n)$, we obtain that $\mathop{\lim}\limits_{n\rightarrow\infty}\frac{\log_2\left|\mathcal{U}^{\prime}(n)\right|}{n}=1$.

\begin{example}\label{exmp_uniquegeneral}
\begin{enumerate}
  \item Let $n=6$. Then $0\le\ell\le\left\lfloor\frac{6}{2}\right\rfloor=3$ and
  \begin{equation*}
    \begin{array}{l}
      \mathcal{P}(6)=\left\{(0,6),(0,1,5),(0,3,3),(0,1,2,4),(0,1,3,3),(0,2,3,3),\right.\\
      \quad\quad\quad\quad\quad\left.(0,1,2,3,3)\right\}.
    \end{array}
  \end{equation*}
  Therefore, $\left|\mathcal{U}^{\prime}(6)\right|=52$.
  \item Let $n=7$. Then $0\le\ell\le\left\lfloor\frac{7}{2}\right\rfloor=3$ and
  \begin{equation*}
    \begin{array}{l}
      \mathcal{P}(7)=\left\{(0,7),(0,1,6),(0,3,4),(0,1,2,5),(0,1,3,4),(0,2,3,4),\right.\\
      \quad\quad\quad\quad\quad\left.(0,1,2,3,4)\right\}.
    \end{array}
  \end{equation*}
  Therefore, $\left|\mathcal{U}^{\prime}(7)\right|=120$.
\end{enumerate}
\end{example}
Comparing Example \ref{exmp_uniquegeneral} with Example \ref{exmp_notunique}, we see that $\cU^{\prime}(6)\ne \cU(6)$, but $\cU^{\prime}(7)= \cU(7)$. 
In general, for $n\geq 6$, it is not clear a priory when $\mathcal{U}^{\prime}(n)= \mathcal{U}(n)$. 
Thus, it is of interest to find a construction for the entire set of uniquely reconstructible (up to reversal) stings. 

\subsection{further generalization}
In this subsection, we build upon the ideas presented in the last two subsections and construct the entire set of uniquely reconstructible (up to reversal) strings.

Consider a string $\bbs\in\Sigma_2^n$ and its generating polynomial $P_{\bbs}(x,y)$ with $\parenv{P_{\bbs}(x,y)}_k=x^{a_k}y^{k-a_k}$. 
For any $1\le i\le\ell+1$, define
\begin{equation*}
\begin{array}{rl}
  P_{\bbs,1}^{(i)}(x,y)&\triangleq x^{-a_{j_{i-1}}}y^{-(j_{i-1}-a_{j_{i-1}})}\mathop{\sum}\limits_{k=j_{i-1}}^{j_i}\left(P_{\bbs}(x,y)\right)_k\\ &=x^{-a_{j_{i-1}}}y^{-(j_{i-1}-a_{j_{i-1}})}\mathop{\sum}\limits_{k=j_{i-1}}^{j_i}x^{a_k}y^{k-a_k},\\
  P_{\bbs,2}^{(i)}(x,y)&\triangleq x^{-w+a_{n-j_{i-1}}}y^{-j_{i-1}+w-a_{n-j_{i-1}}}\mathop{\sum}\limits_{k=j_{i-1}}^{j_i}\left(P_{\bbs}^{*}(x,y)\right)_k\\
  &=x^{-w+a_{n-j_{i-1}}}y^{-j_{i-1}+w-a_{n-j_{i-1}}}\mathop{\sum}\limits_{k=j_{i-1}}^{j_i}x^{w-a_{n-k}}y^{k-(w-a_{n-k})}
\end{array}
\end{equation*}
where $w=a_n=\wth(\bbs)$. Since both $P_{\bbs}(x,y)$ and $P_{\bbs}^{*}(x,y)$ are generating polynomials, we conclude that $P_{\bbs,1}^{(i)}(x,y)$ and $P_{\bbs,2}^{(i)}(x,y)$ are generating polynomials of length-$(j_i-j_{i-1})$ strings. 
Let us denote by $\bm{r}^{(i)}, \bbt^{(i)}\in\Sigma_2^{j_i-j_{i-1}}$, the strings that correspond to $P_{\bbs,1}^{(i)}(x,y), P_{\bbs,2}^{(i)}(x,y)$ respectively. 
In particular, if $i=\ell+1$, then $\bm{r}^{(i)}=\left(\bbt^{(i)}\right)^{*}$. 
From the definition of $j_i$, we obtain that $\left(P_{\bbs}(x,y)\right)_{j_{i-1}}=\left(P_{\bbs}^{*}(x,y)\right)_{j_{i-1}}$, $\left(P_{\bbs}(x,y)\right)_{j_{i}}=\left(P_{\bbs}^{*}(x,y)\right)_{j_{i}}$ and $\left(P_{\bbs}(x,y)\right)_{k}$ $\ne\left(P_{\bbs}^{*}(x,y)\right)_{k}$ for all $j_{i-1}<k<j_{i}$. Therefore, 
\begin{equation}\label{eq_lastgeneralization}
  \begin{array}{l}
    \wth\left(\bm{r}^{(i)}\right)=\wth\left(\bbt^{(i)}\right),\\
    \wth\left(\left(\bm{r}^{(i)}\right)_{1}^{k}\right)\ne \wth\left(\left(\bbt^{(i)}\right)_{1}^{k}\right) \text{ for all } 1\le k<j_i-j_{i-1}.
  \end{array}
\end{equation}
In particular, we have $\bm{r}^{(\ell+1)}\in\mathcal{S}_R(j_{\ell+1}-j_{\ell})\cup\mathcal{S}_R(j_{\ell+1}-j_{\ell})^{*}$. 
Moreover, if $\bm{r}^{(i)}=\left(\bbt^{(i)}\right)^{*}\in\mathcal{S}_R(j_{i}-j_{i-1})\cup\mathcal{S}_R(j_{i}-j_{i-1})^{*}$ for all $1\le i\le \ell+1$, then \Cref{eq_lastgeneralization} holds for all $1\leq i\leq \ell+1$.

Putting everything together, we obtain the following proposition.
\begin{proposition}\label{prop_all}
  For any $n\ge 1$, it holds that
  \begin{equation*}
    \begin{array}{l}
      \Sigma_2^n=\left\{\bm{r}^{(1)}\cdots\bm{r}^{(\ell)}\bm{r}^{(\ell+1)}\left(\bbt^{(\ell)}\right)^{*}\cdots\left(\bbt^{(1)}\right)^{*}:\;\ell\ge 0;j_0=0;\right.\\
      \quad\quad\quad\quad 1\le j_1<\cdots j_{\ell}\le\lfloor\frac{n}{2}\rfloor, j_{\ell+1}=n-j_{\ell};\\
      \quad\quad\quad\quad \bm{r}^{(\ell+1)}\in\mathcal{S}_R(j_{\ell+1}-j_{\ell})\cup\mathcal{S}_R(j_{\ell+1}-j_{\ell})^{*}\\
      \quad\quad\quad\quad\bm{r}^{(i)}, \bbt^{(i)}\in\Sigma_2^{j_i-j_{i-1}}\text{ for all }1\le i\le\ell;\\
      \quad\quad\quad\quad\left.\bm{r}^{(i)}, \bbt^{(i)}\text{ satisfy \Cref{eq_lastgeneralization} for all }1\le i\le\ell\right\}.
    \end{array}
  \end{equation*}
\end{proposition}

Combining Proposition \ref{prop_all} with \Cref{thm_notunique2}, we obtain the following.
\begin{proposition}\label{prop_lastgeneralization}
  For any $n\ge 1$, we have
  \begin{equation*}
    \begin{array}{l}
      \mathcal{U}(n)=\left\{\bm{r}^{(1)}\cdots\bm{r}^{(\ell)}\bm{r}^{(\ell+1)}\left(\bbt^{(\ell)}\right)^{*}\cdots\left(\bbt^{(1)}\right)^{*}:\;\ell\ge 0;j_0=0;\right.\\
      \quad\quad\quad\quad 1\le j_1<\cdots j_{\ell}\le\lfloor\frac{n}{2}\rfloor, j_{\ell+1}=n-j_{\ell};\\
      \quad\quad\quad\quad \bm{r}^{(\ell+1)}\in\mathcal{S}_R(j_{\ell+1}-j_{\ell})\cup\mathcal{S}_R(j_{\ell+1}-j_{\ell})^{*}\\
      \quad\quad\quad\quad\bm{r}^{(i)}, \bbt^{(i)}\in\Sigma_2^{j_i-j_{i-1}}\text{ for all }1\le i\le\ell;\\
      \quad\quad\quad\quad\bm{r}^{(i)}, \bbt^{(i)}\text{ satisfy \Cref{eq_lastgeneralization} for all }1\le i\le\ell;\\
      \quad\quad\quad\quad\left.\text{there is at most one }i\text{ such that }j_i-j_{i-1}\ge2\right\}.
    \end{array}
  \end{equation*}
\end{proposition}
\begin{remark}
  If we take $\bm{r}^{(i)}=\left(\bbt^{(i)}\right)^{*}\in\mathcal{S}_R(j_{i}-j_{i-1})\cup\mathcal{S}_R(j_{i}-j_{i-1})^{*}$ for all $1\le i\le \ell$ we 
  obtain $\mathcal{U}^{\prime}(n)$. Therefore, $\cU(n)$ is indeed a generalization of $\mathcal{U}^{\prime}(n)$.
\end{remark}

Our next aim is to explicitly construct $\bm{r}^{(i)}$ and $\bbt^{(i)}$ so that they satisfy \Cref{eq_lastgeneralization}. To this end, let $\bm{u}=\bm{r}^{(i)}\left(\bbt^{(i)}\right)^{*}$. Then \Cref{eq_lastgeneralization} is equivalent to
\begin{equation}\label{eq_equivlast}
  \begin{array}{l}
    \wth\left(\bm{u}_1^{\frac{m}{2}}\right)=\wth\left(\bm{u}_{\frac{m}{2}+1}^{m}\right),\\
    \wth\left(\bm{u}_1^{k}\right)\ne \wth\left(\bm{u}_{m-k+1}^{m}\right) \text{ for all }1\le k<\frac{m}{2},
  \end{array}
\end{equation}
where $m=2(j_i-j_{i-1})$.

Then the question is reduced to explicitly constructing strings that satisfies \Cref{eq_equivlast}. 
To move further, we first introduce the well known \textit{Dyck strings} \cite[Section 1.4]{gtm290}. A string $\bbs\in\Sigma_2^{2h}$ is called a Dyck string if $\wth(\bbs)=h$ and $\wth(\bbs_1^{i})\ge\lceil\frac{i}{2}\rceil$ for all $1\le i\le 2h-1$. The number of Dyck strings of length $2h$ is given by the Catlan number \cite[Theorem 1.4.1]{gtm290}:
\begin{equation*}
  C_{h}\triangleq\frac{1}{h+1}\binom{2h}{h}.
\end{equation*}
For even $m\ge 4$, let
\begin{equation*}
  \begin{array}{l}
    \mathcal{A}(m)=\left\{\bm{u}\in\Sigma_2^m:\;u_1=1,u_{\frac{m}{2}}=0,u_{\frac{m}{2}+1}=1,u_{m}=0;\right.\\
    \quad\quad\quad\quad\;\;\exists I\subseteq\{2,\ldots,\frac{m}{2}-1\}\text{ such that if }i\ne 1,\frac{m}{2},\text{ and }i\le\frac{m}{2},\\
    \quad\quad\quad\quad\;\;\text{ then }u_i\ne u_{m-i+1}\text{ if and only if }i\in I;\\
    \quad\quad\quad\quad\;\;\left.\bm{u}_{I}\text{ is a Dyck string}\right\}.
  \end{array}
\end{equation*}
By definition, we have
\begin{equation}\label{eq_catlan}
  \left| \mathcal{A}(m)\right|=\mathop{\sum}\limits_{i=0}^{\lfloor\frac{m}{4}-1\rfloor}\binom{\frac{m}{2}-2}{2i}2^{\frac{m}{2}-2-2i}\frac{1}{i+1}\binom{2i}{i}
\end{equation}

If $m=2$, we define $\mathcal{D}(m)\triangleq\{00,11\}$. If $m\ge 4$ is even, we define $\mathcal{D}(m)\triangleq\{\bm{u}:\;\bm{u}\in\mathcal{A}(m)\text{ or }\bm{u}^{*}\in\mathcal{A}(m)\}$.
\begin{lemma}\label{lem_catlan2}
  Let $m\ge 2$ be an even integer. Then a string $\bm{u}\in\Sigma_2^m$ satisfies \Cref{eq_equivlast} if and only if $\bm{u}\in\mathcal{D}(m)$.
\end{lemma}
\begin{IEEEproof}
  The $\Leftarrow$ direction follows immediately from symmetry around $m/2$ in the construction. 
  Thus, we only need to prove the $\Rightarrow$ direction. 
  If $m=2$, the conclusion is trivial so we may assume $m\ge 4$. 
  
  We first notice that since $m\geq 4$, the second row of \Cref{eq_equivlast} has a meaning. From the second row, we deduce $u_1\neq u_m$. 
  Assume that $u_1=1$ and $u_m=0$ (the case $u_1=0$ and $u_m=1$ is similar). 
  Notice that the second row of \Cref{eq_equivlast} implies that $\wth(\bbu_1^k)>\wth(\bbu_{m-k+1}^m)$ for all $1\leq k\leq \frac{m}{2}-1$. 
  Thus, the equality in the first row of \Cref{eq_equivlast}, $\wth(\bbu_1^{m/2})=\wth(\bbu_{m/2+1}^m)$, implies that $u_{m/2}=0$ and $u_{m/2+1}=1$. 
  
  Let $I=\left\{i:\;2\le i\le\frac{m}{2}-1,u_{i}\ne u_{m-i+1}\right\}$. If $I=\emptyset$ we are done. 
  Otherwise, the first row of \Cref{eq_equivlast} implies that $|I|$ is even and $\wth\left(\bm{u}_{I}\right)=\frac{|I|}{2}$. 
  This, together with the second row of \Cref{eq_equivlast} implies that $\bbu_{I}$ is a Dyck string.
\end{IEEEproof}

As a result of Proposition \ref{prop_all} and Lemma \ref{lem_catlan2}, we obtain the following characterization of the set of all binary sequences.
\begin{corollary}\label{cor_all}
For all $n\ge 1$, we have
  \begin{equation*}
  \begin{array}{l}
      \Sigma_2^n=\left\{\bm{r}^{(1)}\cdots\bm{r}^{(\ell)}\bm{r}^{(\ell+1)}\left(\bbt^{(\ell)}\right)^{*}\cdots\left(\bbt^{(1)}\right)^{*}:\;\ell\ge 0;j_0=0\right.\\
      \quad\quad\quad\quad 1\le j_1<\cdots j_{\ell}\le\lfloor\frac{n}{2}\rfloor, j_{\ell+1}=n-j_{\ell};\\
      \quad\quad\quad\quad \bm{r}^{(\ell+1)}\in\mathcal{S}_R(j_{\ell+1}-j_{\ell})\cup\mathcal{S}_R(j_{\ell+1}-j_{\ell})^{*};\\
      \quad\quad\quad\quad\bm{r}^{(i)},\bbt^{(i)}\in\Sigma_2^{j_i-j_{i-1}}\text{ for all }1\le i\le \ell;\\
      \quad\quad\quad\quad\left.\bm{r}^{(i)}\left(\bbt^{(i)}\right)^{*}\in\mathcal{D}(2j_i-2j_{i-1})\text{ for all }1\le i\le\ell\right\}.
    \end{array}
  \end{equation*}
\end{corollary}

\begin{example}
  \begin{enumerate}
    \item Let $n=1$. Then $\ell=0$ and
    \[
    \Sigma_2^n=\mathset{\bbr^{(1)}:\;\bbr^{(1)}\in\cS_R(1)\cup\cS_R(1)^{*}}=\mathset{0,1}.
    \]
    \item Let $n=2$ (then $\ell=1$ or $0$). If $\ell=1$, we have $j_1=j_2=1$ and $\bbr^{(1)},\bbt^{(1)}\in\Sigma_2$. 
    If $\ell=0$, we have $j_1=2$ and $\bm{r}^{(1)}\in\mathcal{S}_R(2)\cup\mathcal{S}_R(2)^{*}$. Therefore,
    \begin{align*}
    \Sigma_2^n&=\left\{\bm{r}^{(1)}\bm{r}^{(2)}\left(\bbt^{(1)}\right)^{*}:\;\bm{r}^{(2)}\in\mathcal{S}_R(0)\cup\mathcal{S}_R(0)^{*}, 
    \bm{r}^{(1)}\left(\bbt^{(1)}\right)^{*}\in\mathcal{D}(2)\right\} 
    \cup\left\{\bm{r}^{(1)}:\;\bm{r}^{(1)}\in\mathcal{S}_R(2)\cup\mathcal{S}_R(2)^{*}\right\}\\
    &=\left\{00,11,01,10\right\},
    \end{align*}
    Since $\mathcal{D}(2)=\left\{00,11\right\}$, $\mathcal{S}_R(0)=\left\{\emptyset\right\}$ and $\mathcal{S}_R(2)=\left\{01\right\}$.
  \end{enumerate}
\end{example}

Combining Proposition \ref{prop_lastgeneralization} and Lemma \ref{lem_catlan2}, we obtain the main result of this subsection.
\begin{theorem}\label{thm_unique}
 For all $n\ge 1$, we have
  \begin{equation*}
    \begin{array}{l}
      \mathcal{U}(n)=\left\{\bm{r}^{(1)}\cdots\bm{r}^{(\ell)}\bm{r}^{(\ell+1)}\left(\bbt^{(\ell)}\right)^{*}\cdots\left(\bbt^{(1)}\right)^{*}:\;\ell\ge 0;j_0=0\right.\\
      \quad\quad\quad\quad 1\le j_1<\cdots j_{\ell}\le\lfloor\frac{n}{2}\rfloor, j_{\ell+1}=n-j_{\ell};\\
      \quad\quad\quad\quad \bm{r}^{(\ell+1)}\in\mathcal{S}_R(j_{\ell+1}-j_{\ell})\cup\mathcal{S}_R(j_{\ell+1}-j_{\ell})^{*}\\
      \quad\quad\quad\quad\bm{r}^{(i)},\bbt^{(i)}\in\Sigma_2^{j_i-j_{i-1}}\text{ for all }1\le i\le \ell;\\
      \quad\quad\quad\quad\bm{r}^{(i)}\left(\bbt^{(i)}\right)^{*}\in\mathcal{D}(2j_i-2j_{i-1})\text{ for all }1\le i\le\ell;\\
      \quad\quad\quad\quad\left.\text{there is at most one }i\text{ such that }j_i-j_{i-1}\ge2\right\}.
    \end{array}
  \end{equation*}
\end{theorem} 

We can also have a formula for the size of $\mathcal{U}(n)$. 
\[
\abs{\cU(n)}=\sum_{\ell=0}^{\floorenv{\frac{n}{2}}} \sum_{\bbj\in\cP(n)}\prod_{i=1}^{\ell+1}g(j_i-j_{i-1}),
\]
where $\bbj$ and $\cP(n)$ are defined in Equation (\ref{eq:P_def}), and 
\begin{equation*}
  g(j_i-j_{i-1})\triangleq
  \begin{cases}
    1, & \mbox{if }j_i=j_{i-1}\\
    2, & \mbox{if }j_i-j_{i-1}=1 \\
    2\left|\mathcal{S}_R(n-2j_{\ell})\right|, & \mbox{if }i=\ell+1\text{ and }n-2j_{\ell}\ge 2\\
    2\left|\mathcal{A}(2j_i-2j_{i-1})\right|, & \mbox{if }j_i-j_{i-1}\ge 2\text{ and }i\ne \ell+1.
  \end{cases}
\end{equation*}

We give two examples below.
\begin{example}\label{examp_uniquemostgeneral}
  Recall that we have obtained $\mathcal{P}(6)$ and $\mathcal{P}(7)$ in Example \ref{exmp_uniquegeneral}.
  \begin{enumerate}
    \item Let $n=6$, we have $\left|\mathcal{U}(6)\right|=60$.
    \item Let $n=7$, we have $\left|\mathcal{U}(7)\right|=120$.
  \end{enumerate}
  The results in this example are consistent with those in Example \ref{exmp_notunique}.
\end{example}

Up to this point we studied the set of strings which can be uniquely reconstructed (up to reversals) from their prefix-suffix composition multiset. 
However, we did not study how many sequences there are that share the same prefix-suffix composition multiset. This is the focus of the next section.

\section{Mutually equivalent strings}\label{sec_equiv}
In this section we focus our attention to the study of sets of strings with a common prefix-suffix composition multiset. 
To clearly state the goals of this subsection, we define the following equivalence relation on the set of strings. 
\begin{definition}
Let $\bbs,\bbt\in\Sigma_2^n$ be two strings of length $n$. We say that $\bbs$ and $\bbt$ are \textit{equivalent}, denoted $\bbs\sim\bbt$, 
if $\cM(\bbs)=\cM(\bbt)$, i.e., if $\bbs$ and $\bbt$ have the same prefix-suffix composition multiset. 
\end{definition}
It is clear that the relation "$\sim$" is indeed an equivalence relation. 
For a string $\bbs\in\Sigma_2^n$ of length $n$, we define
\begin{align*}
  E\parenv{\bbs}\triangleq\mathset{\bbt\in\Sigma_2^n:\;\bbt\sim\bbs}
\end{align*} 
to be the set of all strings that are equivalent to $\bbs$, and 
\begin{align*}
  E_n\triangleq\max\mathset{\abs{E\parenv{\bbs}} ~:~ \bbs\in\Sigma_2^n}.
\end{align*}
The goal of this subsection is to completely characterize the sets $E(\bbs)$ and $E_n$ for any $\bbs$ and $n$. 
To that end, we build upon previous ideas. Specifically, we use similar ideas to those appeared in \Cref{thm_notunique}, \Cref{thm_notunique2}, Proposition \ref{prop_all} and \Cref{thm_unique}.
Before diving into the calculations, we give some simple examples.
\begin{example}
 \begin{enumerate}
   \item Let $n=1$. Then $E\left(0\right)=\left\{0\right\}$, $E\left(1\right)=\left\{1\right\}$ and hence $E_{1}=1$.
   \item Let $n=2$. Then $E\left(00\right)=\left\{00\right\}$, $E\left(11\right)=\left\{11\right\}$, and $E\left(01\right)=\left\{01,10\right\}$. Therefore, $E_{2}=2$.
   \item From Corollary \ref{cor_unique}, we conclude that for $3\le n\le 5$, $\abs{E(\bbs)}=1$ or $2$ for any $\bbs\in\Sigma_2^n$ and that $|E_n|=2$.
 \end{enumerate}
\end{example} 
In the rest of this section, we always assume that $n\ge 2$, unless otherwise is stated. 

Recall that for any $\bbs\in\Sigma_2^n$, we can always find $0=j_0<j_1<\cdots <j_{\ell}\le \lfloor\frac{n}{2}\rfloor$, where $\ell\ge 0$, such that $\left(P_{\bbs}(x,y)\right)_{j}=\left(P_{\bbs}^{*}(x,y)\right)_{j}$ for all $j=j_i,n-j_{i}$, and $\left(P_{\bbs}(x,y)\right)_{j}\neq\left(P_{\bbs}^{*}(x,y)\right)_{j}$ for all $j\ne j_i,n-j_{i}$, where $0\le i\le\ell$. 
In other words, $\wth\left(\bbs_1^j\right)=\wth\left(\bbs_{n-j+1}^n\right)$ for all $j=j_i,n-j_{i}$, and $\wth\left(\bbs_1^j\right)\ne \wth\left(\bbs_{n-j+1}^n\right)$ for all $j\ne j_i,n-j_{i}$. 

Given a string $\bbs$, we define
\begin{equation*}
    I_{\bbs}\triangleq\left\{i:\;1\le i\le\ell+1,j_{i}-j_{i-1}\ge2\right\}.
\end{equation*}
Then \Cref{thm_notunique2} asserts that $\bbs$ cannot be uniquely reconstructed up to reversal if and only if $\left|I_{\bbs}\right|\ge 2$. \Lref{lem_legal} and \Cref{thm_notuniquegeneral} tell us that $\bbt\in E(\bbs)$ if only if is obtained from $\bbs$ by swapping all $\left(P_{\bbs}(x,y)\right)_{j}$'s and $\left(P_{\bbs}^{*}(x,y)\right)_{j}$'s for all $j\in S$, where $S$ is a union of maximal intervals. Therefore, we have the following natural definition.

\begin{definition}
Let $A\subseteq I_{\bbs}$ be a set. We define the set  
\begin{align*}
\swap_1(\bbs,A)=\begin{cases}
  \biggl\{\bbt\in \Sigma_2^n ~:~ \forall i\in A,\; \forall j_{i-1}<j<j_i,\; (P_{\bbt}(x,y))_j=(P_{\bbs}^*(x,y))_j, (P_{\bbt}(x,y))_{n-j}=(P_{\bbs}^*(x,y))_{n-j}, 
  \\ 
  \text{ and } (P_{\bbt}(x,y))_j=(P_{\bbs}(x,y))_j \text{ otherwise } \biggr\}\quad\quad\quad\quad\quad\quad\quad\quad\quad\quad\;\text{ if } A\neq \emptyset\\ 
  \mathset{\bbs}\quad\quad\quad\quad\quad\quad\quad\quad\quad\quad\quad\quad\quad\quad\quad\quad\quad\quad\quad\quad\quad\quad\quad\quad\quad\quad\quad\text{if }A=\emptyset,
\end{cases}
\end{align*} 
and define $\swap(\bbs,A)$ as 
\[\swap(\bbs,A)=\swap_1(\bbs,A)\cup \swap_1(\bbss,A).\] 
In words, $\swap(\bbs,A)$ is the set of all the strings obtained by swapping between $\left(P_{\bbs}(x,y)\right)_j$ and $\left(P_{\bbs}^{*}(x,y)\right)_j$, 
and between $\left(P_{\bbs}(x,y)\right)_{n-j}$ and $\left(P_{\bbs}^{*}(x,y)\right)_{n-j}$, for all $j_{i-1}<j<j_i$ for some $i\in A$, while keeping $\left(P_{\bbs}(x,y)\right)_j$ and $\left(P_{\bbs}^{*}(x,y)\right)_j$ intact for all other $j$'s. 
If $A=\emptyset$, we define $\swap(\bbs,A)=\left\{\bbs,\bbss\right\}$.
\end{definition}

Notice that $|\swap_1(\bbs,A)|=1$ which implies that $1\leq \abs{\swap(\bbs,A)}\leq 2$. Indeed, if $I_{\bbs}=\emptyset$ then $A=\emptyset$. 
This means that $\bbs=\bbss$ and $\swap_1(\bbs,A)=\swap_1(\bbss,A)$ which in turn, implies that $|\swap(\bbs,A)|=1$. 
If $I_{\bbs}\neq \emptyset$, then $\bbs\neq \bbss$ and $\swap_1(\bbs,A)\neq \swap_1(\bbss,A)$, implying that $|\swap(\bbs,A)|=2$. 
\begin{example}
\label{exm_ofirst}
  Let us consider the following sequence 
  \[\bbs=110101011101010111\] 
  with $|\bbs|=18$. 
  We have $\bbss=111010101110101011$. 
  The corresponding polynomials are 
  \begin{align*}
    P_{\bbs}(x,y)&=1+x+x^2+(x^2y)+x^3y+(x^3y^2)+x^4y^2+(x^4y^3)+x^5 y^3+x^6y^3+x^7y^3+(x^7y^4)+x^8y^4+(x^8y^5) \\
    &\quad +x^{9}y^5+(x^{9}y^6)+x^{10}y^6+x^{11}y^6+x^{12}y^6\\
    P_{\bbss}(x,y)&=1+x+x^2+(x^3)+x^3y+(x^4y)+x^4y^2+(x^5 y^2)+ x^5y^3+x^6y^3+x^7y^3+(x^8y^3)+x^8y^4 +(x^9y^4)\\ 
    &\quad +x^{9}y^5+(x^{10}y^5)+x^{10}y^6+x^{11}y^6+x^{12}y^6,
\end{align*} 
where $\parenv{\cdot}$ denotes a difference in the total degree between $P_{\bbs}(x,y)$ and $P_{\bbss}(x,y)$. 
Thus, we obtain $\floorenv{18/2}=9$ and $j_1=1,j_2=2,j_3=4,j_4=6,j_5=8,j_6=9$ with the additional $j_{\ell+1}=j_7=18-j_6=9$. 
This implies that $I_{\bbs}=\mathset{3,4,5}$ since $j_3-j_2=j_4-j_3,j_5-j_4=2$. 
Let us take $A=\mathset{3,4}$. 
The set $\swap_1(\bbs,A)$ contains a string $\bbt_1$ for which 
\begin{align*}
    P_{\bbt_1}(x,y)&=1+x+x^2+(x^3)+x^3y+(x^4y)+x^4y^2+x^4y^3+x^5 y^3+x^6y^3+x^7y^3+x^7y^4+x^8y^4+(x^9y^4) \\
    &\quad +x^{9}y^5+(x^{10}y^5)+x^{10}y^6+x^{11}y^6+x^{12}y^6. 
\end{align*} 
Thus, the corresponding string is $\bbt_1=111010011101101011$. 
Similarly, $\swap_1(\bbss,A)$ contains a string $\bbt_2$ for which 
\begin{align*}
    P_{\bbt_2}(x,y)&=1+x+x^2+(x^2y)+x^3y+(x^3y^2)+x^4y^2+x^5y^2+x^5 y^3+x^6y^3+x^7y^3+x^8y^3+x^8y^4+(x^8y^5) \\
    &\quad +x^{9}y^5+(x^{9}y^6)+x^{10}y^6+x^{11}y^6+x^{12}y^6. 
\end{align*} 
Thus, the corresponding string is $\bbt_2=110101101110010111=\bbt_1^{\star}$. 
Overall, $\swap(\bbs,A)=\mathset{\bbt_1,\bbt_2}=\mathset{\bbt_1,\bbt_1^{\star}}$.
\end{example}

Recall that in \Cref{sec_notunique}, we have proved that $\bbt\sim\bbs$ if and only if $\bbt\in \swap(\bbs,A)$ for some $A\subseteq I_{\bbs}$.
\begin{lemma}\label{lem_disjoint}
  Let $A,B$ be two subsets of $I_{\bbs}$ such that $A\neq B$. 
  \renewcommand{\labelenumi}{(\roman{enumi})}
  \begin{enumerate}
      \item If $A\cup B\ne I_{\bbs}$, then $\swap(\bbs,A)\cap \swap(\bbs,B)=\emptyset$.
      \item If $A\cup B= I_{\bbs}$ but $A\cap B\ne\emptyset$, then $\swap(\bbs,A)\cap \swap(\bbs,B)=\emptyset$.
      \item If $A\cup B= I_{\bbs}$ and $A\cap B=\emptyset$, then $\swap(\bbs,A)= \swap(\bbs,B)$.
  \end{enumerate}
\end{lemma}

\begin{IEEEproof}
(\Rnum{1})
 Since $A\cup B\ne I_{\bbs}$,
there exists $i_0\in I_{\bbs}\setminus \left(A\cup B\right)$ which implies that there is $j_{i_0-1}<j<j_{i_0}$ for which  $\left(P_{\bbs}(x,y)\right)_{j}\ne\left(P_{\bbs}^{*}(x,y)\right)_{j}$. 
  By definition, $\swap(\bbs,A)=\swap_1(\bbs,A)\cup\swap_1(\bbss,A)$ and $\swap(\bbs,B)=\swap_1(\bbs,B)\cup\swap_1(\bbss,B)$. 
  Assume $\swap(\bbs,A)\cap\swap(\bbs,B)\neq \emptyset$, i.e., there exists $\bbt\in \swap(\bbs,A)\cap\swap(\bbs,B)$. 
  For $\bbt$, we have 
  \[\parenv{P_{\bbt}(x,y)}_{j}\in\mathset{\parenv{P_{\bbs}(x,y)}_{j},\parenv{P_{\bbs}^*(x,y)}_{j}}.\] 
  Assume $\parenv{P_{\bbt}(x,y)}_{j}=\parenv{P_{\bbs}(x,y)}_{j}$. 
  This implies that $\bbt\in\swap_1(\bbs,A)\cap \swap_1(\bbs,B)$, which, in turn, implies that $\parenv{P_{\bbs}(x,y)}_{j^{\prime}}=\parenv{P_{\bbs}^*(x,y)}_{j^{\prime}}$ 
  for all $j_{i-1}<j^{\prime}<j_i$ such that $i\in A\triangle B$ where $\triangle$ is the symmetric difference. 
  This is possible only if $A\triangle B=\emptyset$ which contradicts the assumption. 
  A similar contradiction arises if we assume $\parenv{P_{\bbt}(x,y)}_{j}=\parenv{P_{\bbs}^*(x,y)}_{j}$. 
  
(\Rnum{2})
Let $C=A\cap B$. Since $C\ne\emptyset$, we can choose an $i_1\in C$. Then  $\parenv{P_{\bbs}(x,y)}_{j^{\prime}}\ne\parenv{P_{\bbs}^*(x,y)}_{j^{\prime}}$ for all $j_{i_1-1}<j^{\prime}<j_{i_1}$. Since $A\ne B$, we have $A\setminus C\ne\emptyset$ or $B\setminus C\ne\emptyset$.

Assume that $A\setminus C\ne\emptyset$ and choose $i_2\in A\setminus C=A\setminus B$. 
We have $\parenv{P_{\bbs}(x,y)}_{j^{\prime\prime}}\ne\parenv{P_{\bbs}^*(x,y)}_{j^{\prime\prime}}$ for all $j_{i_2-1}<j^{\prime\prime}<j_{i_2}$. 
Let $\bbt_1\in\swap(\bbs,A)$ and $\bbt_2\in\swap(\bbs,B)$ be arbitrary. 
For $\bbt_1$, we have either 
\[\parenv{P_{\bbt_1}(x,y)}_{j^{\prime}}=\parenv{P_{\bbs}(x,y)}_{j^{\prime}},\;  \parenv{P_{\bbt_1}(x,y)}_{j^{\prime\prime}}=\parenv{P_{\bbs}(x,y)}_{j^{\prime\prime}},\] 
or 
\[\parenv{P_{\bbt_1}(x,y)}_{j^{\prime}}=\parenv{P_{\bbs}^*(x,y)}_{j^{\prime}},\;  \parenv{P_{\bbt_1}(x,y)}_{j^{\prime\prime}}=\parenv{P_{\bbs}^*(x,y)}_{j^{\prime\prime}}.\]
On the other hand, for $\bbt_2$, we have either 
\[\parenv{P_{\bbt_2}(x,y)}_{j^{\prime}}=\parenv{P_{\bbs}(x,y)}_{j^{\prime}},\; \parenv{P_{\bbt_2}(x,y)}_{j^{\prime\prime}}=\parenv{P_{\bbs}^*(x,y)}_{j^{\prime\prime}},\] 
or 
\[\parenv{P_{\bbt_2}(x,y)}_{j^{\prime}}=\parenv{P_{\bbs}^*(x,y)}_{j^{\prime}},\;  \parenv{P_{\bbt_2}(x,y)}_{j^{\prime\prime}}=\parenv{P_{\bbs}(x,y)}_{j^{\prime\prime}}.\] 
This implies that $\bbt_1\ne\bbt_2$ and since $\bbt_1,\bbt_2$ were arbitrary, $\swap(\bbs,A)\cap \swap(\bbs,B)=\emptyset$. 
Similar analysis holds when $B\setminus C\ne\emptyset$.

(\Rnum{3})
By definition, it follows that $\swap_1(\bbs,A)=\swap_1(\bbss,B)$ and $\swap_1(\bbss,A)=\swap_1(\bbs,B)$. Therefore, $\swap(\bbs,A)=\swap(\bbs,B)$.
\end{IEEEproof}

Let us continue Example \ref{exm_ofirst}. 
\begin{example}
  Consider the same string $\bbs=110101011101010111$ as in Example \ref{exm_ofirst}, with the same polynomials $P_{\bbs}(x,y)$ and $P_{\bbs}^*(x,y)$. 
  Let us consider the case $B=\mathset{4,5}$. 
  The set $\swap_1(\bbs,B)$ contains a string $\bbw_1$ for which 
  \begin{align*}
    P_{\bbw_1}(x,y)&=1+x+x^2+x^2y+x^3y+(x^4y)+x^4y^2+(x^5 y^2)+ x^5y^3+x^6y^3+x^7y^3+(x^8y^3)+x^8y^4 +(x^9y^4)\\ 
    &\quad +x^{9}y^5+x^{9}y^6+x^{10}y^6+x^{11}y^6+x^{12}y^6. 
\end{align*} 
The corresponding string in $\bbw_1=110110101110100111$. 
Similarly, the set $\swap_1(\bbss,B)$ contains a string $\bbw_2$ for which 
\begin{align*}
    P_{\bbw_1}(x,y)&=1+x+x^2+x^3+x^3y+(x^3y^2)+x^4y^2+(x^4y^3)+x^5 y^3+x^6y^3+x^7y^3+(x^7y^4)+x^8y^4+(x^8y^5) \\
    &\quad +x^{9}y^5+x^{10}y^5+x^{10}y^6+x^{11}y^6+x^{12}y^6. 
\end{align*}
The corresponding string in $\bbw_2=111001011101011011=\bbw_1^{\star}$. 

Notice that taking $B'=\mathset{5}$ yields $\swap(\bbs,B')=\swap(\bbs,A)$ and taking $A'=\mathset{3}$ in Example \ref{exm_ofirst} yields $\swap(\bbs,B)=\swap(\bbs,A')$.
\end{example}

Using the previous lemma, we can characterize the set of strings that are equivalent to $\bbs$, i.e., the set $E(\bbs)$.
To that end, we define the following. 
\begin{definition}
Let $S$ be a set and let $A$ be a set of subsets of $S$, i.e., $A\subseteq 2^S$ where $2^S$ is the power set of $S$. 
We say that $A$ is \textit{open under complements} if $X\in A$ implies $X^c=S\setminus X\notin A$. 
If $A\subseteq 2^S$ we denote by $\ouc{A}$ the set obtained by making $A$ an open under complements set by arbitrarily 
eliminating $X$ or $X^c$ if both $X,X^c\in A$, i.e., if $X\in A$ and $S\setminus X\in A$ we choose one of them arbitrarily and remove it from $A$.
\end{definition}

For a string $\bbs$, let $A=2^{I_{\bbs}}$ be the power set of $I_{\bbs}$. 
For simplicity of notation, we define $\cF_{\bbs}=\ouc{A}$, meaning that $\cF_{\bbs}$ is an open under complement set obtained from the power set of $I_{\bbs}$. 
Of course $\mathcal{F}_{\bbs}$ is not unique.
For example, if $I_{\bbs}=\{1,4\}$, then the power set is $2^{I_{\bbs}}=\mathset{\emptyset,\{1\},\{4\},\{1,4\}}$. 
We can take $\mathcal{F}_{\bbs}=\mathset{\emptyset,\{1\}},\; \mathcal{F}_{\bbs}=\mathset{\emptyset,\{4\}}, \mathcal{F}_{\bbs}=\mathset{\{1\},\{1,4\}}$, or 
$\mathcal{F}_{\bbs}=\mathset{\{4\},\{1,4\}}$. 
However, if $I_{\bbs}\ne\emptyset$, it is easy to see that $\left|\mathcal{F}_{\bbs}\right|=2^{\left|I_{\bbs}\right|-1}$, 
regardless of the choice of $\mathcal{F}_{\bbs}$.

\begin{corollary}\label{cor_num2}
  For any $\mathcal{F}_{\bbs}$, we have $E\left({\bbs}\right)=\bigcup\limits_{A\in \mathcal{F}_{\bbs}}\swap(\bbs,A)$. In particular, $\left|E\left({\bbs}\right)\right|=2^{\left|I_{\bbs}\right|}$.
\end{corollary}
\begin{IEEEproof}
  The first equality follows from noticing that $\cM(\bbs)=\cM(\bbt)$ for every $\bbt\in\swap(\bbs,A)$ for some $A\subseteq I_{\bbs}$ and from  Lemma \ref{lem_disjoint}. Indeed, Lemma \ref{lem_disjoint} asserts that if $A\cup B=I_{\bbs}$ and $A\cap B=\emptyset$ then $\swap(\bbs,A)=\swap(\bbs,B)$. 
  Thus, instead of going over all $A\subseteq I_{\bbs}$, it suffices to consider sets $A$ without their complement. 
  Thus, we are left to prove the second equality.

  If $I_{\bbs}=\emptyset$, then $\bbs=\bbss\in\mathcal{U}(n)$, implying $\left|E\left({\bbs}\right)\right|=1=2^{\left|I_{\bbs}\right|}$. 
  Assume that $I_{\bbs}\ne\emptyset$, i.e., $\left|\swap(\bbs,A)\right|=2$ for each $A\subseteq I_{\bbs}$. 
  In other words, each $A\subseteq I_{\bbs}$ contributes $2$ to $\left|E\left({\bbs}\right)\right|$. 
  On the other hand, from the construction of $\cF_{\bbs}$ we have that for $A,B\in \cF_{\bbs}$, either $A\cup B\ne I_{\bbs}$ or $A\ne I_{\bbs}\setminus B$. 
  In either case, Lemma \ref{lem_disjoint} asserts that $\swap(\bbs,A)\cap \swap(\bbs,B)=\emptyset$. 
  Therefore, $\left|E\left({\bbs}\right)\right|=2\left|\mathcal{F}_{\bbs}\right|=2^{\left|I_{\bbs}\right|}$.
\end{IEEEproof}

We present some examples to explain Corollary \ref{cor_num2}.
\begin{example}\label{examp_swap}
  \begin{enumerate}
    \item Let $n=6$ and $\bbs=101010$. From Example \ref{examp_partition}, we already know that $\ell=1$, $j_1=2$, and $j_{\ell+1}=j_2=4$. 
    Therefore, $I_{\bbs}=\left\{1,2\right\}$ since $j_2-j_1=j_1-j_0=2$. 
    Let $\mathcal{F}_{\bbs}=\left\{\emptyset,\{1\}\right\}$. By definition, we have $\swap(\bbs,\emptyset)=\left\{\bbs,\bbss\right\}$ and $\swap(\bbs,\{1\})=\left\{011001,100110\right\}$. Then by Corollary \ref{cor_num2}, we have
        $$
        E\left({\bbs}\right)=\swap(\bbs,\emptyset)\cup \swap(\bbs,\{1\})=\left\{101010,010101,011001,100110\right\}
        $$
        and $\left|E\left({\bbs}\right)\right|=2^{\left|I_{\bbs}\right|}=2^{2}=4$.
    \item Let $n=6$ and $\bbs=010111$. From Example \ref{examp_partition}, we know that $\ell=0$ so $I_{\bbs}=\{1\}$. 
    By Corollary \ref{cor_num2}, we have
        $
        E\left({\bbs}\right)=\swap(\bbs,\emptyset)=\left\{\bbs,\bbss\right\}
        $
        and $\left|E\left({\bbs}\right)\right|=2^{\left|I_{\bbs}\right|}=2$.
    \item Let $n=7$ and $\bbs=1011010$. From Example \ref{examp_partition}, we know that $\ell=1$ and $j_1=2$, 
    so $I_{\bbs}=\{1,2\}$. 
    By definition, we have $\swap(\bbs,\emptyset)=\left\{\bbs,\bbss\right\}$ and $\swap(\bbs,\{1\})=\left\{0111001,1001110\right\}$. 
    Then by Corollary \ref{cor_num2}, we have
        $$
        E\left({\bbs}\right)=\swap(\bbs,\emptyset)\cup \swap(\bbs,\{1\})=\left\{1011010,0101101,0111001,1001110\right\}
        $$
        and $\left|E\left({\bbs}\right)\right|=2^{\left|I_{\bbs}\right|}=4$.
  \end{enumerate}
\end{example}
We see that the results in Example \ref{examp_swap} are consistent with those in Example \ref{exmp_notunique}.

Our final goal for this section is to estimate the maximal size of $E(\bbs)$, i.e., to calculate $E_n$. 
From Corollary \ref{cor_num2}, we see that in order to determine $E_n$, we should determine the maximum value of $\left|I_{\bbs}\right|$. 
\begin{theorem}\label{thm_maxnum}
  For $n\ge 2$, we have $E_n= 2^{\floorenv{\frac{n-2}{4}}+1}$.
\end{theorem}
\begin{IEEEproof} 
The proof is a case-by-case analysis.
\begin{enumerate}
    \item Assume $n$ is even. 
    \begin{description}
      \item[If $j_{\ell}=\frac{n}{2}$:] We have $j_{\ell+1}-j_{\ell}=n-2j_{\ell}=0$. Then $|I_{\bbs}|\leq \floorenv{\frac{n/2}{2}}=\floorenv{\frac{n}{4}}$.
      \item[If $j_{\ell}=\frac{n}{2}-1$:] We have $j_{\ell+1}-j_{\ell}=2$. 
      Then $\left|I_{\bbs}\right|\le \floorenv{\frac{\frac{n}{2}-1}{2}}+1=\floorenv{\frac{n-2}{4}}+1$. 
          This bound can be achieved by taking $j_{i}-j_{i-1}=2$ for all $1\le i\le\ell+1$ with possibly one exception in which $j_{i}-j_{i-1}=1$.
      \item[If $j_{\ell}\le \frac{n}{2}-2$:] We have $j_{\ell+1}-j_{\ell}\ge 4$. 
      Thus, $\left|I_{\bbs}\right|\le \floorenv{\frac{\frac{n}{2}-2}{2}}+1=\floorenv{\frac{n}{4}}$. 
    \end{description}
          Thus, if $n$ is even, then $E_n=\max\left\{2^{\floorenv{\frac{n}{4}}}, 2^{\floorenv{\frac{n-2}{4}}+1}\right\}=2^{\floorenv{\frac{n-2}{4}}+1}$. 
    \item Assume $n$ is odd. 
    \begin{description}
      \item[If $j_{\ell}=\floorenv{\frac{n}{2}}$:] We have $j_{\ell+1}-j_{\ell}=1$. 
      Then $|I_{\bbs}|\le \floorenv{\frac{\floorenv{\frac{n}{2}}}{2}}= \floorenv{\frac{n-1}{4}}$.
      \item[If $j_{\ell}\le\floorenv{\frac{n}{2}}-1$:] We have $j_{\ell+1}-j_{\ell}\ge 3$. 
      Thus, $|I_{\bbs}|\le \floorenv{\frac{\floorenv{\frac{n}{2}}-1}{2}}+1=\floorenv{\frac{n-3}{4}}+1=\floorenv{\frac{n-2}{4}}+1$. 
      This bound can be achieved by taking $j_{\ell}=\floorenv{\frac{n}{2}}-1$ and $j_{i}-j_{i-1}=2$ for all $1\le i\le\ell$ with possibly one exception in which $j_{i}-j_{i-1}=1$.
    \end{description} 
    Therefore, if $n$ is odd, then $E_n=\max\mathset{2^{\floorenv{\frac{n-1}{4}}},2^{\floorenv{\frac{n-2}{4}}+1}}=2^{\floorenv{\frac{n-2}{4}}+1}$.
\end{enumerate}
Overall, we obtain that $E_n=2^{\floorenv{\frac{n-2}{4}}+1}$ which is the wanted result.
\end{IEEEproof}

\begin{remark}
  Notice that Theorem \ref{thm_maxnum} asserts that $E_n\ge 4>2$ for all $n\ge 6$. 
  This means that there exists a string $\bbs$ of length $n$ that cannot be uniquely reconstructed up to reversal for every $n\geq 6$.
\end{remark}

Up to now we studied strings that can (or cannot) be uniquely reconstructed up to reversal. 
Although the condition of unique reconstruction up to reversal makes sense from a computational point of view, for application purposes, we prefer strings which can be uniquely reconstructed (not up to reversal). This is the main focus of the next section.

\section{Composition codes}\label{sec_codes}
In this section we study strings which can be uniquely reconstructed from their composition multiset. To that end, we define a \textit{composition code}, 
$\mathcal{C}\subseteq\Sigma_2^n$, as a set of strings such that for $\bbs,\bbt\in\cC$, $\mathcal{M}(\bbs)= \mathcal{M}(\bbt)$ implies $\bbs=\bbt$. 
That is to say, any $\bbs\in\mathcal{C}$ can be uniquely reconstructed. 
For example, consider the set $\cC_0$ that contains the strings $\bbs$ such that $\bbs=\bbss$, or $\bbs\in\cC_0$ implies $\bbss\notin\cC_0$, i.e., 
\begin{equation*}
  \mathcal{C}_0=\left\{\bbs\in\mathcal{U}(n):\;\bbs=\bbss,\text{ or exactly one of }\bbss\text{ and }\bbs\in\mathcal{C}_0\right\}.
\end{equation*}
If follows from Theorem \ref{thm_unique} that $\mathcal{C}_0$ is a composition code.

Before we present our construction, we mention some relevant results on composition codes. 
The authors of \cite{wz2} studied the composition multiset reconstruction in the general case for which several strings are to be reconstructed simultaneously. 
In this work, we consider the case of reconstructing only one string. Thus, it is reasonable to assume that the construction presented in \cite{wz2} can be improved in the case of a single string reconstruction. 
Indeed, using the construction in \cite[Theorem 10]{wz2}, we get a composition code $C(n,1)$ of size whose order of magnitude is $2^{n-\sqrt{n}}$, where $n=m+\frac{17}{2}\sqrt{m}$ for some positive integer $m$. 
Considering the code $\cC_0$ above, we have $|\cC_0|=\frac{|\cU(n)|-|\cC_1(n)|}{2}+|\cC_1(n)|=\frac{|\cU(n)|+|\cC_1(n)|}{2}$ where 
$\cC_1(n)=\mathset{\bbs\in\cU(n) ~:~ \bbs=\bbss}$.  
Hence $\cC_0$ is of order of magnitude of at least $\frac{2^{n-3}}{\sqrt{n}}$ by Equation (\ref{eq_sizeofS}). 
Thus, for the specific case of a single string reconstruction, in terms of code size, $\cC_0$ improves upon the construction given in \cite[Theorem 10]{wz2}. 
Notice, however, that $\cC_0$ is not optimal in terms of code size since we can add strings $\bbs\notin \cU(n)$ for $n\geq 6$. 

Our goals in this section is to study optimal composition codes (in terms of code size). 
In the next subsections we present several constructions for composition codes, some of which are capable of correcting errors. 
We start with constructions under the assumptions that there are no errors while reading the compositions. 

\subsection{Error-free prefix-suffix composition multiset}
In this subsection we present two constructions of composition codes. 
We first notice that by definition, a subset $\mathcal{C}\subseteq\Sigma_2^n$ is a composition code if and only if $|\cC\cap E(\bbs)|\le 1$ for any $\bbs\in\Sigma_2^n$. 
Therefore, the following code construction is optimal.
\begin{construction}\label{cons_maxcode}
  For all $n\ge 6$, let $\cC_{max}\subseteq\Sigma_2^n$ be such that $|\cC\cap E(\bbs)|= 1$ for any $\bbs\in\Sigma_2^n$.
\end{construction}

We can use Corollary \ref{cor_num2} to explicitly determine $E(\bbs)$ for any $\bbs\in\Sigma_2^n$. 
Therefore, Construction \ref{cons_maxcode} is indeed an explicit construction. 
Note that the code $\mathcal{C}_{max}$ is not unique in general.

Although any given $\mathcal{C}_{max}$ is an optimal code in terms of code size, 
we do not know how to uniquely reconstruct an arbitrary $\bbs\in\mathcal{C}_{max}$ when given $\mathcal{M}(\bbs)$. 
Fortunately, it is possible to construct a code such that every string $\bbs$ in that code can be uniquely reconstructed from $\cM(\bbs)$. 

For a string $\bbs\in\Sigma^n$, let $\mathcal{M}_p(\bbs),\cM_s(\bbs)$ denote the multiset of all the compositions of prefixes of $\bbs$, and the multiset of 
all suffixes of $\bbs$, respectively. 
When given $\mathcal{M}_p(\bbs)$ or $\mathcal{M}_s(\bbs)$, it is straightforward to (uniquely) reconstruct $\bbs$, 
so the idea in the next construction is to impose additional constraints on $\bbs$ such that when given $\mathcal{M}(\bbs)$, 
we can distinguish $\mathcal{M}_p\left(\bbs\right)$ from $\mathcal{M}_s\left(\bbs\right)$. 
With this idea and Corollary \ref{cor_all} in mind, it is now natural to give the following construction.

\begin{construction}\label{cons_errorfree1}
  For all $n\ge 6$, let
  \begin{equation*}
  \begin{array}{l}
      \mathcal{E}_1(n)=\left\{\bm{r}^{(1)}\cdots\bm{r}^{(\ell)}\bm{r}^{(\ell+1)}\left(\bbt^{(\ell)}\right)^{*}\cdots\left(\bbt^{(1)}\right)^{*}:\;\ell\ge 0;j_0=0\right.\\
      \quad\quad\quad\quad 1\le j_1<\cdots <j_{\ell}\le\left\lfloor\frac{n}{2}\right\rfloor, j_{\ell+1}=n-j_{\ell};\\
      \quad\quad\quad\quad \bm{r}^{(\ell+1)}\in\mathcal{S}_R(j_{\ell+1}-j_{\ell})^{*};\\
      \quad\quad\quad\quad\bm{r}^{(i)},\bbt^{(i)}\in\Sigma_2^{j_i-j_{i-1}}\text{ for all }1\le i\le \ell;\\
      \quad\quad\quad\quad\left.\bm{r}^{(i)}\left(\bbt^{(i)}\right)^{*}\in\mathcal{A}(2j_i-2j_{i-1})\text{ for all }1\le i\le\ell\right\}.
    \end{array}
  \end{equation*}
\end{construction}
It follows from the construction of $\cE_1(n)$ that for every $\bbs\in\mathcal{E}_1(n)$, $\wth\left(\bbs_1^j\right)\ge \wth\left(\bbs_{n-j+1}^n\right)$ 
for all $1\le j\le n$. 
Recall that $\bbt\sim\bbs$ if and only if $\bbt\in \swap(\bbs,A)$ for some $A\subseteq I_{\bbs}$. 
It is straightforward to verify that if $\bbs\in\mathcal{E}_1(n)$, $\bbt\sim\bbs$, and $\bbt\ne\bbs$, then there exists some $j_0$ such that $\wth\left(\bbt_1^{j_0}\right)< \wth\left(\bbt_{n-j_0+1}^n\right)$ and hence $\bbt\notin\mathcal{E}_1(n)$. 
That is to say, $\mathcal{E}_1(n)$ is a composition code. 
Furthermore, since $\wth\left(\bbs_1^j\right)\ge \wth\left(\bbs_{n-j+1}^n\right)$ for all $1\le j\le n$, it is possible to distinguish $\mathcal{M}_p(\bbs)$ 
from $\mathcal{M}_s\left(\bbs\right)$ when given $\mathcal{M}\left(\bbs\right)$. 
Thus, any $\bbs\in\mathcal{E}_1(n)$ can be uniquely reconstructed from $\mathcal{M}\left(\bbs\right)$. 
The reconstruction process is straightforward by the discussions above. 
The size of $\mathcal{E}_1(n)$ is
\begin{equation}\label{eq_errorfree1}
\left|\mathcal{E}_1(n)\right|=\sum_{\ell=0}^{\lfloor\frac{n}{2}\rfloor}\sum_{\bbj\in\mathcal{P}^{\prime}(n)}\prod_{i=1}^{\ell+1} h(j_i-j_{i-1}),
\end{equation}
where $j_0=0$,  
$$
\mathcal{P}^{\prime}(n)\triangleq\left\{\bbj=(j_1,\ldots,j_{\ell+1}) ~:~ \ell\ge 0,1\le j_1<\cdots j_{\ell}\le\left\lfloor\frac{n}{2}\right\rfloor, j_{\ell+1}=n-j_{\ell}\right\},
$$ 
and
\begin{equation*}
  h(j_i-j_{i-1})\triangleq
  \begin{cases}
    1, & \mbox{if }j_i=j_{i-1}\\
    2, & \mbox{if }j_i-j_{i-1}=1 \\
    \left|\mathcal{S}_R(n-2j_{\ell})\right|, & \mbox{if }i=\ell+1\text{ and }n-2j_{\ell}\ge 2\\
    \left|\mathcal{A}(2j_i-2j_{i-1})\right|, & \mbox{if }j_i-j_{i-1}\ge 2\text{ and }i\ne \ell+1.
  \end{cases}
\end{equation*}
By letting $\ell=0$, it follows from \Cref{eq_errorfree1} that $\left|\mathcal{E}_1(n)\right|>\left|\mathcal{S}_R(n)\right|\ge\frac{2^{n-3}}{\sqrt{\pi n}}$. 
Hence, the redundancy of $\mathcal{E}_1(n)$ is less than $\frac{1}{2}\log_2(n)+3+\frac{1}{2}\log_2(\pi)$ and the code $\mathcal{E}_1(n)$ has asymptotic rate $1$.

It is worth noticing that Construction \ref{cons_errorfree1} is under the assumption that the prefix-suffix composition 
multiset $\mathcal{M}\left(\bbs\right)$ is error-free. 
Unfortunately, this assumption is false in practical applications because the MS/MS readout technique is error prone. 
The authors of \cite{wz2} considered the following two types of errors.
\begin{enumerate}
  \item \textit{Missing of prefixes or suffixes}. These type of errors occur when the masses of some prefixes and suffixes are not measured and/or reported.
  \item \textit{Mass reducing substitution errors}. These type of errors occur when a composition is mistakenly read out as a composition of the same length, 
  but with a smaller Hamming weight, i.e., a larger mass is reported as a smaller mass due to loss of atoms during the fragmentation process.
\end{enumerate}
More details regarding the errors can be found in \cite[Section \RNum{5}]{wz2}. The following are examples for the type of errors mentioned above. 
\begin{example}\label{examp_errortype}
  Let $\bbs=101010$. One can check that $\bbs=101010\in \mathcal{E}_1(6)$ with $\ell=1$. 
  Consider the erroneous prefix-suffix composition multiset
  $$
  \widetilde{\mathcal{M}}_1\left(\bbs\right)=\left\{1,01,01^2,0^21^2,0^31^3,0^31^3,0^31^2,0^21^2,0^21,01,0\right\},
  $$
  one can  see that the prefix composition $0^21^3$ is missing. Now consider the erroneous prefix-suffix composition multiset
  $$
  \widetilde{\mathcal{M}}_2\left(\bbs\right)=\left\{1,01,01^2,0^21^2,0^31^2,0^31^3,0^31^3,0^31^2,0^21^2,0^21,01,0\right\},
  $$
  one can see that the prefix composition $0^21^3$ is mistakenly read out as $0^31^2$. 
  Now let $\bbt=100110$ and consider the erroneous prefix-suffix composition multiset
  $$
  \widetilde{\mathcal{M}}_3\left(\bbt\right)=\left\{1,01,0^21,0^21^2,0^31^2,0^31^3,0^31^3,0^31^2,0^21^2,01^2,01,0\right\},
  $$
  one can see that the prefix composition $0^21^3$ is mistakenly read out as $0^31^2$. This makes $\widetilde{\cM}_3(\bbt)=\widetilde{\cM}_2(\bbs)$. 
\end{example}

In the rest of this section, we will modify Construction \ref{cons_errorfree1} to obtain composition codes that can correct these two type of errors.

\subsection{Missing of prefixes or suffixes}
In this subsection, we aim to construct composition codes capable of correcting the first type of errors. 
Suppose that there are at most $t$ prefix compositions or suffix compositions are missing. 
For a given $\bbs\in\mathcal{E}_1(n)$, even if the prefix composition of length $i$ is missing but the suffix composition of length $i$ remains unchanged 
(or vice versa), we will still regard both of them are missing, due to the fact that we cannot identify whether the remained one is the prefix composition 
or a suffix. 
Therefore, this error corresponds to at most $2t$ erasures in the reconstructed string $\widetilde{\bbs}$ (see \cite{wz2}). 
With this in mind, we can encode strings in $\mathcal{E}_1(n)$ by the following three steps:
\begin{description}
  \item[Step 1:] Choose a systematic $[m,n,\ge 2t+1]$ code $\mathcal{E}$ over finite field $\mathbb{F}_q$, where $q=2^s$ is a power of 2. Fix a basis $\mathcal{B}$ of $\mathbb{F}_q$ over $\mathbb{F}_2$.
  \item[Step 2:] For each $\bbs\in\mathcal{E}_1(n)$, we firstly encode it using $\mathcal{E}$ into $\bbs\bm{r}^{\prime}$, where $\bm{r}^{\prime}$ is the string of redundancy bits. Secondly, according to $\mathcal{B}$, we convert each symbol of $\bm{r}^{\prime}$ into a binary string of length $s$. Then we concatenate these length-$s$ binary strings into $\bm{r}$. At last, we get an intermediary string $\bbs\bm{r}$.
  \item[Step 3:] The output codeword is $\bm{c}=\bm{r}^{*}\bbs\bm{r}\in\Sigma_2^N$, where $N=n+2sr(\cE)$ and $r(\cE)$ is the redundancy of $\mathcal{E}$. 
  This step is to ensure that $\wth\left(\bm{c}_1^j\right)\ge \wth\left(\bm{c}_{N-j+1}^N\right)$ for all $1\le j\le N$.
\end{description}
The decoding of $\bbs$ from $\bm{r}^{*}\bbs\bm{r}$ is straightforward. Denote the resulting code by $\mathcal{E}_2(n)$. Then $r\left(\mathcal{E}_2(n)\right)=r\left(\mathcal{E}_1(n)\right)+2s\cdot r\left(\mathcal{E}\right)$.

Since for each cyclic code there is a systematic encoder \cite[Page 128]{huffman_pless_2003} and BCH codes are in most cases among the best cyclic codes, we can let $\mathcal{E}$ be a narrow-sense primitive BCH code over $\mathbb{F}_2$ with designed distance $2t+1$, then $r(\mathcal{E})\le t\log_2(m+1)$ \cite[Theorem 5.1.7]{huffman_pless_2003}. Then $m-t\log_2(m+1)\le n\le m-2t$ and $m-t\log_2(m+1)+4t\le N\le m-2t+2t\log_2(m+1)$. Now we have
$$
\frac{\log_2\left(\left|\mathcal{E}_2(n)\right|\right)}{N}=\frac{\log_2\left(\left|\mathcal{E}_1(n)\right|\right)}{N}>\frac{m-t\log_2(m+1)-3-\log_2(\sqrt{\pi})-\log_2(\sqrt{m-2t})}{m-2t+2t\log_2(m+1)}.
$$
Therefore, the code $\mathcal{E}_2(n)$ has asymptotic rate $1$.

\subsection{Mass reducing substitution errors}
In this subsection, we aim to construct composition codes capable of correcting the second type of errors. 
Throughout this part, we assume that errors are not in batches, i.e., if the composition of the prefix length $i$ is erroneous, then the composition of the prefixes lengths $i-1$ and $i+1$ are read out correctly, and similarly, if the composition of the suffix length $i$ is erroneous, then the composition of the suffixes length $i-1$ and $i+1$ are read out correctly. 
Although this requirement seems kind of impractical, we believe that the ideas we used to obtain the results in this subsection 
will shed some light into further study of this type of errors.

First, we study the different effects an error has on a string. 
Assume  $0^{a_1}1^{b_1}$, $0^{a_2}1^{b_2}$, and $0^{a_3}1^{b_3}$ are the compositions of prefixes of length $i-1$, $i$ and $i+1$, respectively, of $\bbs$, 
where $1\le i\le n$. 
Here we define $a_1=b_1=0$ if $i=0$, and $a_3=a_2+1$, $b_3=b_2$ if $i=n$, thus $b_1\le b_2\le b_1+1$ and $b_3-1\le b_2\le b_3$. 
Suppose that the composition $0^{a_2}1^{b_2}$ is mistakenly read out as $0^{a_2^{\prime}}1^{b_2^{\prime}}$, where $b_2^{\prime}<b_2$ and $a_2^{\prime}+b_2^{\prime}=a_2+b_2=i$. 
Furthermore, suppose that the two compositions $0^{a_1}1^{b_1}$ and $0^{a_3}1^{b_3}$ are error free. 
Assume also that given the erroneous prefix-suffix composition multiset $\widetilde{M}\left(\bbs\right)$, we can still distinguish $\widetilde{M}_p\left(\bbs\right)$ from $\widetilde{M}_s\left(\bbs\right)$ (this requirement will be fulfilled by the presented code). 

Note that $b_1\le b_3\le b_1+2$. 
If $b_3=b_1$ or $b_3=b_1+2$, then the error can be detected and corrected directly. 
If $b_3=b_1+1=b_2+1$ or $b_3=b_2=b_1+1$, and $b_2-b_2^{\prime}\ge 2$, then the error can be detected. 
If $b_3=b_2=b_1+1$ and $b_2-b_2^{\prime}=1$, then $b_1\le b_2^{\prime}\le b_1+1$ and $b_3-1\le b_2^{\prime}\le b_3$. 
In this case, we cannot detect the error because $\bbs_1^{i+1}=\bbs_1^{i-1}10$ will be reconstructed as $\bbs_1^{i-1}01$ 
(recall that we define $s_{n+1}=0$ if $i=n$).

According to the above analyses, we divide the mass reducing substitution errors into two types. 
If $b_3=b_2=b_1+1$ and $b_2-b_2^{\prime}=1$, we say this error is \textit{compatible}. For instance, the prefix-suffix composition multiset $\widetilde{\mathcal{M}}_3(\bbt)$ in Example \ref{examp_errortype} is obtained by introducing a compatible error.
If $b_3=b_2=b_1+1$ but $b_2-b_2^{\prime}\ge 2$, we say it is \textit{incompatible}. 
Note that a single compatible error will result in two adjacent substitution errors in the reconstructed string, 
while a single incompatible error will result in two adjacent erasures in the reconstructed string.

We now define the error model. For given integers $e_1,e_2\ge 1$ and $t\ge 2$, we formulate an $(e_1,e_2,t)$-mass reducing substitution error model as follows:
\begin{itemize}
  \item If a composition $0^a 1^b$ is mistakenly read out as $0^{a+b-b^{\prime}}1^{b^{\prime}}$, then $1\le b-b^{\prime}\le t$;
  \item There are at most $e_1$ compatible errors;
  \item There are at most $e_2$ incompatible errors.
  \item If the composition of the prefix length $i$ is erroneous, then the composition of the prefixes length $i-1$ and $i+1$ are not erroneous;
  \item If the composition of the suffix length $i$ is erroneous, then the composition of the suffixes length $i-1$ and $i+1$ are not erroneous.
\end{itemize}

We can encode $\mathcal{E}_1(n)$ as follows:
\begin{description}
  \item[Step 1:] Choose a systematic $[m,n,d]$ code $\mathcal{E}$ over finite field $\mathbb{F}_q$ that can correct $2e_1$ substitutions and $2e_2$ erasures, where $q=2^s$ is a power of 2. Fix a basis $\mathcal{B}$ of $\mathbb{F}_q$ over $\mathbb{F}_2$.
  \item[Step 2:] For each $\bbs\in\mathcal{E}_1(n)$, we firstly encode it using $\mathcal{E}$ into $\bbs\bm{r}^{\prime}$, where $\bm{r}^{\prime}$ is the string of redundancy bits. Secondly, according to $\mathcal{B}$, we convert each symbol of $\bm{r}^{\prime}$ into a binary string of length $s$. Then we concatenate these length-$s$ binary string into $\bm{r}$. At last, we get an intermediary string $\bbs^{\prime}=\bm{r}^{*}\bbs\bm{r}\in\Sigma_2^{n_1}$. This step is to ensure that $\wth\left(\left(\bbs^{\prime}\right)_1^j\right)\ge \wth\left(\left(\bbs^{\prime}\right)_{n_1-j+1}^{n_1}\right)$ for all $1\le j\le n_1$.
  \item[Step 3:] The output codeword is $\bm{c}=\bm{1}^{t}\bm{r}^{*}\bbs\bm{r}\bm{0}^{t}\in\Sigma_2^N$, where $N=n+2sr\left(\mathcal{E}\right)$ and $r\left(\mathcal{E}\right)$ is the redundancy of $\mathcal{E}$. This step is to ensure that $\wth\left(\bm{c}_1^j\right)\ge \wth\left(\bm{c}_{N-j+1}^{N}\right)+t$ for all $t+1\le j\le N-t$.
\end{description}
Denote the resulting code by $\mathcal{E}_3(n)$. Then $r\left(\mathcal{E}_3(n)\right)=r\left(\mathcal{E}_1(n)\right)+2s\cdot r\left(\mathcal{E}\right)+2t$.

Since the first $t$ bits of $\bm{c}$ are all $1$'s and the last $t$ bits of $\bm{c}$ are all $0$'s, we can distinguish $\widetilde{M}_p\left(\bbs\right)$ from $\widetilde{M}_s\left(\bbs\right)$, due to the property that $\wth\left(\bm{c}_1^j\right)\ge \wth\left(\bm{c}_{N-j+1}^{N}\right)+t$ for all $t+1\le j\le N-t$. Now the decoding of $\bbs$ from $\bm{1}^{t}\bm{r}^{*}\bbs\bm{r}\bm{0}^{t}$ is straightforward. Therefore, the code $\mathcal{E}_3(n)$ can correct $(e_1,e_2,t)$-mass reducing substitution errors.

\begin{lemma}\cite[Theorem 1.11.6]{huffman_pless_2003}\label{lem_erasure-error}
  Let $\mathcal{C}$ be an $[m,n,d]$ code, Then $\mathcal{C}$ is capable of
correcting any pattern of $2e_1$ substitution errors and $2e_1$ erasures if $4e_1+2e_2 \le d-1$.
\end{lemma}

With Lemma \ref{lem_erasure-error} in hand, we can let $\mathcal{E}$ be a narrow-sense primitive BCH code over $\mathbb{F}_2$ with designed distance $2\left(2e_1+e_2\right)+1$, then $r(\mathcal{E})\le \left(2e_1+e_2\right)\log_2(m+1)$ \cite[Theorem 5.1.7]{huffman_pless_2003}. Then $m-\left(2e_1+e_2\right)\log_2(m+1)\le n\le m-2\left(2e_1+e_2\right)$ and $N\le m-2\left(2e_1+e_2\right)+2\left(2e_1+e_2\right)\log_2(m+1)+2t$. Now we have
$$
\begin{array}{rl}
\frac{\log_2\left(\left|\mathcal{E}_3(n)\right|\right)}{N}&=\frac{\log_2\left(\left|\mathcal{E}_1(n)\right|\right)}{N}\\
&>\frac{m-\left(2e_1+e_2\right)\log_2(m+1)-3-\log_2(\sqrt{\pi})-\log_2\left(\sqrt{m-2\left(2e_1+e_2\right)}\right)}{m-2t+2t\log_2(m+1)}.
\end{array}
$$
Therefore, the code $\mathcal{E}_3(n)$ has asymptotic rate $1$.

\section{Deleted compositions from the entire composition multiset}\label{sec_delcomp}
In this section, we turn our attention to the second subclass of the general problem: reconstructing a string from its composition multiset when at most a constant number of substring compositions are lost. Notice that in this section we consider the problem of string reconstruction from the entire composition multiset and not only from the prefix-suffix composition multiset. 

Let us denote by $C(\bbs)$ the composition multiset of a string $\bbs\in\Sigma_2^n$, i.e., $C(\bbs)=\left\{c(\bbs_i^j)~:~1\le i\le j\le n\right\}$. For each $1\le l\le n$, let $C_l(\bbs)$ be the multiset of compositions of \underline{all} the substrings of length $l$, i.e., $C_l(\bbs)=\left\{c\left(\bbs_i^{i+l-1}\right):\:1\le i\le n-l+1\right\}$. 
Then we have $C(\bbs)=\mathop{\cup}\limits_{l=1}^{n}C_l(\bbs)$. 
Our problem can be formally stated as follows.

\textbf{Problem : } Let $L\subseteq \mathset{1,2,\dots,n}$ be a set of size $t_1$ and assume that for every $l\in L$, $C_l(\bbs)$ may suffer the loss of at most $t_2$ compositions. Here we assume that $t_1$ and $t_2$ are both constants. Is there a code which is resilient to that error model? 
In other words, a code such that each codeword can be reconstructed even if such an error happens?

If $\mathcal{C}$ is such a code, we say $\mathcal{C}$ is a \textit{$(t_1,t_2)$-deletion correcting composition code}. 
We also say $\mathcal{C}$ can correct a $(t_1,t_2)$-deletion error. 
By definition, $\mathcal{C}$ is a $(t_1,t_2)$-deletion correcting composition code if and only if for any two distinct strings $\bbs,\bbv\in\cC$, 
there is no subset $\mathcal{I}\subseteq\{1,\ldots,n\}$ of size at most $t_1$ such that
\begin{itemize}
  \item $C_l(\bbs)\ne C_l(\bbv)$ if and only if $l\in\mathcal{I}$;
  \item $\left|C_l(\bbs)\setminus C_l(\bbv)\right|\le t_2$ for each $l\in\mathcal{I}$.
\end{itemize}

The authors of \cite{wz17} studied stronger cases in which a finite number (say $t_1$) of $\mathcal{C}_l$'s are totally missing, and not just a part of their elements. 
Their codes are naturally $(t_1,t_2)$-deletion correcting composition codes for any $t_2$. 
Since the construction of \cite{wz17} is resilient to any $t_2$, it is reasonable that their codes can be improved 
(in terms of size) for a limited value of $t_2$. 
We improve upon the construction of \cite{wz17}, and present a $(t_1,t_2)$-deletion correcting compositions codes of larger size for the case $t_2=2$. 
In the rest of this section, we always assume that $n$ is an even integer. 

For a string $\bbs\in\Sigma_2^n$, define $w_l(\bbs)\triangleq \sum\limits_{i=1}^{n-l+1}\wth\left(\bbs_i^{i+l-1}\right)$ for any $1\le l\le n$. 
For each $1\le i\le\frac{n}{2}$, denote $\sigma_i(\bbs)=\wth(s_i s_{n-i+1})$, i.e., the number of ones in the pair of symbols $s_i,s_{n-i+1}$. 
Finally, let $\bm{\sigma}(\bbs)\triangleq\left(\sigma_1(\bbs),\ldots,\sigma_{\frac{n}{2}}(\bbs)\right)$. 
It is straightforward to show that $w_l(\bbs)=w_{n-l+1}(\bbs)$ for any $l$. 
It was proved in \cite[Section II-A]{wz3} that given $w_1(\bbs),\dots,w_{\frac{n}{2}}(\bbs)$, 
it is possible to uniquely determine $\bm{\sigma}(\bbs)$.
For an integer $t\ge 1$, a set $\mathcal{S}^{(t)}_R (n)$ was constructed in \cite[Section IV-B]{wz3} as follows:
\begin{equation*}
  \begin{array}{rl}
    \mathcal{S}_R^{(t)}(n)&=\left\{\bbs\in\Sigma_2^n:\;s_1^{t}=\0,s_{n-t+1}^n=\1,\text{ and }\exists I\subset\{t+1,\ldots,n-t\}\text{ such that when }t+1\le i\le n-t,\right.\\
    &\quad\;\;\text{we have }s_i\ne s_{n-i+1}\text{ for all }i\in I\text{ and }s_i=s_{n-i+1}\text{ for all }i\notin I;\\
    &\quad\;\;\left.\bbs_{\left[\frac{n}{2}\right]\cap I}\text{ is a Catalan-Bertrand string}\right\}.
  \end{array}
\end{equation*}
Before we connect $\cS_R^{(t)}(n)$ to $(t_1,t_2)$-deletion correcting composition codes, we need the following definition. 
We say that $i$ is a \textit{prefix-suffix pair} shared by two $n$-length strings $\bbs,\bbv$ if $\bbs_1^i=\bbv_1^i$, $\bbs_{n-i+1}^n=\bbv_{n-i+1}^n$. 
The \textit{longest prefix-suffix pair} shared by $\bbs$ and $\bbv$ is the largest $i$ such that $i$ is a prefix-suffix pair and $i+1$ is not, i.e., 
if $\bbs_1^i=\bbv_1^i$ and $\bbs_{n-i+1}^n=\bbv_{n-i+1}^n$ but $s_{i+1}\ne v_{i+1}$ or $s_{n-i}\ne v_{n-i}$.
The following lemma was obtained in \cite{wz3}. 
\begin{lemma}\cite[Lemma 4]{wz3}\label{lem_SRt}
  Let $\bbs,\bbv\in\mathcal{S}_R^{(t)}(n)$ be two distinct strings with $\bm{\sigma}(\bbs)=\bm{\sigma}(\bbv)$. 
  If the longest prefix-suffix pair shared by $\bbs$ and $\bbv$ is of length $i$, 
  then $\left|C_{n-i-k}(\bbs)\setminus C_{n-i-k}(\bbv)\right|\ge 2$ for all $1\le k\le t+1$. 
\end{lemma}
By taking $t=t_1$, we obtain that $\mathcal{S}_R^{(t_1)}(n)$ is a $(t_1,t_2)$-deletion correcting composition code for any $t_2\ge 1$. 
Comparing the definition of $(t_1,t_2)$-deletion correcting composition codes and Lemma \ref{lem_SRt}, we see that $\mathcal{S}_R^{(t_1)}(n)$ has a stronger-than-need property, that is, $C_l(\bbs)\ne C_l(\bbv)$ for at least $t_1+1$ such $l$'s. 
Therefore, to increase the code size, we aim at adding strings to $\mathcal{S}_R^{(t_1)}(n)$. 
To this end, recall the definition of the set $\mathcal{S}_R(n)$ given in Equation (\ref{eq_srn}). 
Clearly, $\mathcal{S}_R^{(t)}(n)\subsetneq\mathcal{S}_R(n)$ for all $t\ge 1$, so we try to pick suitable strings in 
$\mathcal{S}_R(n)\setminus \mathcal{S}_R^{(t)}(n)$ and add them to $\mathcal{S}_R^{(t)}(n)$.

Let $\bbs,\bbv\in\mathcal{S}_R(n)$ be two distinct strings and assume $\bm{\sigma}(\bbs)=\bm{\sigma}(\bbv)=(\sigma_1,\ldots,\sigma_{\frac{n}{2}})$. 
When $\sigma_i\in\{0,2\}$, we have $s_i=s_{n-i+1}=v_{i}=v_{n-i+1}$. 
As a result, if $\sigma_i\in\{0,2\}$ for all $i\ge 2$, then $\bbs=\bbv$, which is a contradiction. 
Thus, there exists some $i\ge 2$ such that $\sigma_i=1$ and denote by $i_1$ the smallest one. 
By the definition of $\mathcal{S}_R(n)$, we obtain that $s_1s_{i_1}$ and $v_1v_{i_1}$ are  both Catalan-Bertrand strings and hence $s_1s_{i_1}=00=v_1v_{i_1}$ (recall that in a Catalan-Bertrand string the number of zeros is strictly larger than the number of ones in any prefix).
So there exists some $i>i_1$ such that $\sigma_i=1$. Otherwise, we have $\bbs=\bbv$, which is a contradiction. Now we can assume the indices $2\le i_1<i_2<\cdots<i_l\le\frac{n}{2}$ being such that $\sigma_i=1$ if and only if $i\in\{1,i_1,i_2,\ldots,i_l\}$, where $l\ge2$.

From above analysis, we know that $s_{i_1}=v_{i_1}$ which, in turn, implies that if $s_i\ne v_i$ then $i\in\{i_2,\ldots,i_l\}$. 
Let $k\ge2$ be the smallest integer such that $s_{i_k}\ne v_{i_k}$. Without loss of generality, we assume $s_{i_k}=0$ and $v_{i_k}=1$.
\begin{lemma}\label{lem_ge3}
   With the above notations, we have $\left|C_{n-i_k-1}(\bbs)\setminus C_{n-i_k-1}(\bbv)\right|\ge 3$ 
   if and only if one of the following conditions is satisfied:
  \renewcommand{\labelenumi}{(\roman{enumi})}
 \begin{enumerate}
 \item $k<l$, $i_{k+1}-i_k=1$ and $(s_{i_{k+1}},v_{i_{k+1}})\ne(0,0),(1,0)$;
 \item $k<l$, $i_{k+1}-i_k\ge2$;
 \item  $k=l<\frac{n}{2}$.
 \end{enumerate}
\end{lemma}
\begin{IEEEproof}
In the following we estimate $\Delta\triangleq\left|C_{n-i_{k}-1}(\bbs)\setminus C_{n-i_{k}-1}(\bbv)\right|$. 
To ease notation, let $\bbu=s_{i_1}\cdots s_{i_{k-1}}=v_{i_1}\cdots v_{i_{k-1}}$. 
Let $N_0 \triangleq N_0(\bbu) =\left|\left\{j:\;s_{i_j}=0,1\le j\le k-1\right\}\right|$, i.e., $N_0$ is the number of zeros in $\bbu$. 
Besides, let $N_1$ $\triangleq$ $N_1(\bbu)=k-1-N_0(\bbu)$ .
Since $v_1v_{i_1}\cdots v_{i_{k-1}}v_k$ is a Catalan-Bertrand string, $v_1=0$ and $v_{i_k}=1$, we have that the number of zeros in $\bbu$ is strictly larger than the number of ones in $\bbu$, i.e., $N_0(\bbu)\ge N_1(\bbu)+1$.
Since we have $s_j=s_{n-j+1}=v_j=v_{n-j+1}$ for every $j\notin\mathset{1,i_1,\dots,i_l}$, and since $s_{i_j}=v_{i_j}$ for $1\leq j<k$, 
it follows that the possible elements in $C_{n-i_{k}-1}(\bbs)$ that are not in $C_{n-i_{k}-1}(\bbv)$ are 
$$
L=\left\{c\left(\bbs_{1}^{n-i_{k}-1}\right),c\left(\bbs_{2}^{n-i_{k}}\right),c\left(\bbs_{i_k+1}^{n-1}\right),c\left(\bbs_{i_k+2}^{n}\right)\right\}.
$$
Similarly, the possible elements in $C_{n-i_{k}-1}(\bbv)$ that are not in $C_{n-i_{k}-1}(\bbs)$ are
$$
R=\left\{c\left(\bbv_{1}^{n-i_{k}-1}\right),c\left(\bbv_{2}^{n-i_{k}}\right),c\left(\bbv_{i_k+1}^{n-1}\right),c\left(\bbv_{i_k+2}^{n}\right)\right\}.
$$

We first consider the case $2\le k<l$. In this case, we have 
\begin{equation*}
  \begin{array}{llllllllrlll}
     &  &  & i_k &  &  &  &  &  & & & \\
   \bbs= s_1 \cdots & s_{i_1} \cdots & s_{i_2} \cdots & 0 \cdots & s_{i_{k+1}} \cdots & s_{i_l} \cdots & \cdots \overline{s_{i_l}}& \cdots \overline{s_{i_{k+1}}}&\cdots 1&\cdots \overline{s_{i_2}}&\cdots \overline{s_{i_1}} &\cdots s_n,  \\
   \bbv= v_1 \cdots & v_{i_1} \cdots & v_{i_2} \cdots & 1 \cdots & v_{i_{k+1}} \cdots & v_{i_l} \cdots & \cdots \overline{v_{i_l}}& \cdots \overline{v_{i_{k+1}}}&\cdots 0&\cdots \overline{v_{i_2}}&\cdots \overline{v_{i_1}} &\cdots v_n
  \end{array}
\end{equation*}
where $\overline{s_{j}}=1-s_j$ is the complement of $s_j$. 
Our aim is to compare between elements in $L$ and elements in $R$. Therefore, we may disregard common entries that appear simultaneously in all the elements that in $L$ and in $R$. As a result, if $i_{k+1}-i_k=1$, we only need to consider the compositions of 
\begin{align*}
    &s_1\bbu s_{i_k}s_{i_{k+1}}s_{i_{k+2}}\cdots \overline{s}_{i_{k+2}}, \\
    &\bbu s_{i_k}s_{i_{k+1}}s_{i_{k+2}}\cdots \overline{s}_{i_{k+2}}\overline{s}_{i_{k+1}},\\
    &s_{i_{k+2}}\cdots \overline{s}_{i_{k+2}}\overline{s}_{i_{k+1}}\overline{s}_{i_k}\overline{\bbu}s_n,\\
    &s_{i_{k+1}}\cdots \overline{s}_{i_{k+1}}\overline{s}_{i_k}\overline{\bbu},
\end{align*}
and 
\begin{align*}
    &v_1\bbu v_{i_k}v_{i_{k+1}}v_{i_{k+2}}\cdots \overline{v}_{i_{k+2}},\\
    &\bbu v_{i_k}v_{i_{k+1}}v_{i_{k+2}}\cdots \overline{v}_{i_{k+2}}\overline{v}_{i_{k+1}},\\ 
    &v_{i_{k+2}}\cdots \overline{v}_{i_{k+2}}\overline{v}_{i_{k+1}}\overline{v}_{i_k}\overline{\bbu}v_n,\\
    &v_{i_{k+1}}\cdots \overline{v}_{i_{k+1}}\overline{v}_{i_k}\overline{\bbu}.
\end{align*}
Thus, we redefine the sets $L,R$ as follows:
\begin{align*}
L=\left\{c\left(\bbu\right)0^2s_{i_{k+1}},c\left(\bbu\right)0^21,c\left(\overline{\bbu}\right)1^2\overline{s}_{i_{k+1}},c\left(\overline{\bbu}\right)01^2\right\},\\
R=\left\{c\left(\bbu\right)01v_{i_{k+1}},c\left(\bbu\right)01^2,c\left(\overline{\bbu}\right)01\overline{v}_{i_{k+1}},c\left(\overline{\bbu}\right)0^21\right\}.
\end{align*}
Notice that in the redefined sets $L,R$ we eliminated $s_{i_{k+2}},s_{i_{k+3}},\dots$ and $v_{i_{k+2}},v_{i_{k+3}},\dots$. This is because for every $i_{k+1}<j\leq i_l$, the symbols $s_j,\bar{s_j}$ appear in $\bbs$ and contribute a single zero and a single one to the composition. Similarly, $v_j,\bar{v_j}$ appear in $\bbv$ and contribute a single zero and a single one as well. 
Thus, both $s_{i_{k+2}},s_{i_{k+3}},\dots$ and $v_{i_{k+2}},v_{i_{k+3}},\dots$ can be eliminated from the modified sets $L,R$.
According to the values of $s_{i_{k+1}}$ and $v_{i_{k+1}}$, we have four subcases:
\begin{itemize}
  \item If $s_{i_{k+1}}=v_{i_{k+1}}=0$, we have $c\left(\bbu\right)0^21$, $c\left(\overline{\bbu}\right)01^2$ $\in L\cap R$ and so $\Delta=|L\setminus R|\le 2$.
  \item If $s_{i_{k+1}}=v_{i_{k+1}}=1$, we have
  \begin{align*}
  L=\left\{c\left(\bbu\right)0^21,c\left(\bbu\right)0^21,c\left(\overline{\bbu}\right)01^2,c\left(\overline{\bbu}\right)01^2\right\},\\
  R=\left\{c\left(\bbu\right)01^2,c\left(\bbu\right)01^2,c\left(\overline{\bbu}\right)0^21,c\left(\overline{\bbu}\right)0^21\right\}.
  \end{align*}
  It is clear that $c\left(\bbu\right)0^21\ne c\left(\bbu\right)01^2$ and $c\left(\overline{\bbu}\right)01^2\ne c\left(\overline{\bbu}\right)0^21$. On the other hand, if $c\left(\bbu\right)0^21=c\left(\overline{\bbu}\right)0^21$, then $N_0(\bbu)=N_0(\overline{\bbu})=N_1(\bbu)$, which contradicts the fact that $N_0(\bbu)\ge N_1(\bbu)+1$.
  Therefore, $L\cap R=\emptyset$ and $\Delta=|L\setminus R|=4$.
  \item If $s_{i_{k+1}}=0,v_{i_{k+1}}=1$, with similar arguments as above, we have $\Delta=|L\setminus R|=4$.
  \item If $s_{i_{k+1}}=1,v_{i_{k+1}}=0$, then $\Delta=|L\setminus R|\le 2$ since $c\left(\bbu\right)0^21$, $c\left(\overline{\bbu}\right)01^2$ $\in L\cap R$.
\end{itemize}
Similarly, if $i_{k+1}-i_k\ge2$, we can redefine $L$ and $R$ as follows:
\begin{align*}
L=\left\{c\left(\bm{u}\right)0^2,c\left(\bm{u}\right)0\delta,c\left(\bar{\bm{u}}\right)1^2,c\left(\bar{\bm{u}}\right)1\delta\right\},\\
R=\left\{c\left(\bm{u}\right)01,c\left(\bm{u}\right)1\delta,c\left(\bar{\bm{u}}\right)01,c\left(\bar{\bm{u}}\right)0\delta\right\},
\end{align*}
where $\delta=s_{i_k+1}=s_{n-i_k}$. For each $\delta\in\{0,1\}$, we have $\Delta=|L\setminus R|= 3$.

Next we consider the case $k=l$. In this case, we have
\begin{equation*}
  \begin{array}{llllllll}
     &  &  &i_k &  &  &  & \\
   \bbs= s_1 \cdots & s_{i_1} \cdots & s_{i_{l-1}} \cdots & 0 \cdots &\cdots 1& \cdots \overline{s_{i_{l-1}}}&\cdots \overline{s_{i_1}} &\cdots s_n,  \\
   \bbv= v_1 \cdots & v_{i_1} \cdots & v_{i_{l-1}} \cdots & 1 \cdots &\cdots 0& \cdots \overline{v_{i_{l-1}}}&\cdots \overline{v_{i_1}} &\cdots v_n.
  \end{array}
\end{equation*}
If $i_l=\frac{n}{2}$, we have $\Delta=|L\setminus R|\le 2$. So we assume that $i_l<\frac{n}{2}$. In this case, $L$ and $R$ can be redefined as
$$
L=\left\{c\left(\bm{u}\right)0^2,c\left(\bm{u}\right)0\delta,c\left(\bar{\bm{u}}\right)1^2,c\left(\bar{\bm{u}}\right)1\delta\right\}
$$
and
$$
R=\left\{c\left(\bm{u}\right)01,c\left(\bm{u}\right)1\delta,c\left(\bar{\bm{u}}\right)01,c\left(\bar{\bm{u}}\right)0\delta\right\},
$$
where $\delta=s_{i_l+1}=s_{n-i_l}$. Similar to the case $k<l$ and $i_{k+1}-i_k\ge2$, we have $\Delta=|L\setminus R|= 3$.
\end{IEEEproof}

For each constant $t$ with $2\le t<\frac{n}{2}-4$, let
$$
\mathcal{I}^{(t)}=\left\{I=\{i_1,i_2,i_3\ldots,i_l\}:\;l\ge3,i_1\ge t+1,i_l<\frac{n}{2},i_{k+1}-i_k\ge 2\text{ for all }2\le k<l\right\}.
$$
Recall the definition of $\cS_R(n)$ (see Equation (\ref{eq_srn})) and let
$$
S^{(t)}=\left\{\bbs\in\mathcal{S}_R(n) ~:~ I=\emptyset,\text{ or }I=\{i_1\},i_1\ge t+1,\text{ or }I=\{i_1,i_2\},t+1\le i_1< i_2<\frac{n}{2},\text{ or }I\in\mathcal{I}\right\}.
$$
By construction, it follows that $S^{(t)}\subseteq\mathcal{S}_R(n)\setminus\mathcal{S}_R^{(t)}(n)$.
\begin{theorem}
  Let $\mathcal{C}^{(t)}_A=S^{(t)}\cup\mathcal{S}_R^{(t)}(n)$. Then $\mathcal{C}^{(t)}_A$ can correct a $(t,2)$-deletion error as long as for any two erroneous $C_k$ and $C_l$, we have $l\ne n-k+1$.\footnote{The requirement that at most one of $C_l$ and $C_{n-l+1}$ is erroneous for each $l$ was previously studied under the name 
  \textit{asymmetric error}. Two kinds of such asymmetric errors have been studied before. Specifically, the authors of \cite{wz3} studied the error model where at most one of $C_l$ and $C_{n-l+1}$ suffered substitution error; the authors of \cite{wz17}  studied the error model where at most one of $C_l$ and $C_{n-l+1}$ was totally deleted.}
\end{theorem}
\begin{IEEEproof}
Let $\bbs\in\mathcal{C}^{(t)}_A$ be a codestring and we are given $\widetilde{C}(\bbs)$ which is obtained from $C(\bbs)$ by introducing a $(t,2)$-deletion error.
Since for any two erroneous $C_k(\bbs)$ and $C_l(\bbs)$ we have $l\ne n-k+1$, the value of $w_l(\bbs)$ is known for every $l$ and thus $\sigma_l(\bbs)$ ($1\le l\le\frac{n}{2}$) can be obtained.

By the constructions of $S^{(t)}$ and $\mathcal{S}_R^{(t)}(n)$, we can see that if $\bbs\in S^{(t)}$ and $\bbv\in\mathcal{S}_R^{(t)}(n)$, then $\bm{\sigma}(\bbs)\ne\bm{\sigma}(\bbv)$. So if two strings $\bbs,\bbv\in\mathcal{C}_A^{(t)}$ satisfies $\bm{\sigma}(\bbs)=\bm{\sigma}(\bbv)$, then $\bbs,\bbv\in S^{(t)}$ or $\bbs,\bbv\in\mathcal{S}_R^{(t)}(n)$. If $\bbs,\bbv\in S^{(t)}$, there is at least one $l$ such that $\left|C_l(\bbs)\setminus C_l(\bbv)\right|\ge3$ by Lemma \ref{lem_ge3}. If $\bbs,\bbv\in\mathcal{S}_R^{(t)}(n)$, there are at least $3$ $l$'s such that $C_l(\bbs)\ne C_l(\bbv)$ by Lemma \ref{lem_SRt}.

Therefore, with $\widetilde{C}(\bbs)$, and $\sigma_l(\bbs)$ ($1\le l\le\frac{n}{2}$), we can perform the Backtracking Algorithm (see \cite[Example 1]{wz3} for details). 
to uniquely reconstruct $\bbs$. Specifically, for step $i$, if $\sigma_i(\bbs)=0$ (or $2$), then $s_i=s_{n-i+1}=0$ (or $1$). If $\sigma_i(\bbs)=1$, there are two possibilities: $(s_i,s_{n-i+1})=(0,1)$ and $(s_i,s_{n-i+1})=(1,0)$. So we need to make a guess and continue to recover $s_{i+1}$ and $s_{n-i}$. The discussion in the second paragraph ensures that when we make a wrong guess, we will derive a string $\bbv$ such that $C(\bbv)$ is incompatible with $\widetilde{C}(\bbs)$.
\end{IEEEproof}

Let $M_0=1$ and $M_i$ denote the number of Catlan-Bertrand strings of length $i$, where $i\ge1$.
From \cite[Theorem 6]{wz3}, we can see that $M_i=\binom{i-1}{\frac{i-1}{2}}$ if $i$ is odd, and $M_i=\frac{1}{2}\binom{i}{\frac{i}{2}}$ if $i\ge 2$ is even. Then \cite[Proposition 1]{wz3} implies that there are at least $\frac{2^{i-1}}{\sqrt{\pi(\left\lfloor\frac{i}{2}\right\rfloor+1)}}$ Catlan-Bertrand strings of length $i$. So by the definition of $\mathcal{S}_R^{(2)}(n)$, we have
\begin{align}\label{eq_srt}
\left|\mathcal{S}_R^{(t)}(n)\right|=\sum_{i=0}^{\frac{n}{2}-t}\binom{\frac{n}{2}-t}{i}2^{\frac{n}{2}-t-i}M_i\ge\sum_{i=0}^{\frac{n}{2}-t}\binom{\frac{n}{2}-t}{i}2^{\frac{n}{2}-t-i}\frac{2^{i-1}}{\sqrt{\pi(\left\lfloor\frac{i}{2}\right\rfloor+1)}}\ge\frac{2^{n-2t-1}}{\sqrt{\pi(\left\lfloor\frac{n}{4}-\frac{t}{2}\right\rfloor+1)}}.
\end{align}
Therefore, 
\begin{equation*}
\log_2\left(\mathcal{C}^{(t)}_A\right)<\log_2\left(\left|\mathcal{S}_R^{(t)}(n)\right|\right)\le
\begin{cases}
  \frac{1}{2}\log_2(n-2t+4)+2t+\frac{1}{2}\log_2(\pi),\mbox{ if } \frac{n}{2}-t \mbox{ is even},\\
  \frac{1}{2}\log_2(n-2t+2)+2t+\frac{1}{2}\log_2(\pi),\mbox{ if } \frac{n}{2}-t  \mbox{ is odd}.
\end{cases}
\end{equation*}
Comparing this upper bound with \cite[Lemma 6]{wz17}, in the suggested setting, it is possible to obtain a code with larger size.

For completeness, in the rest of this section we derive the size of $\mathcal{C}_A^{(t)}$. 
Since $S^{(t)}\subseteq\mathcal{S}_R(n)\setminus\mathcal{S}_R^{(t)}(n)$, the key point is to enumerate the strings in $S^{(t)}$. 
Since $M_1=M_2=1$ and $M_3=2$, there are $2^{\frac{n}{2}-1}$ strings $\bbs\in S^{(t)}$ 
with $I=\emptyset$, $\left(\frac{n}{2}-t\right)2^{\frac{n}{2}-2}$ strings 
$\bbs\in S^{(t)}$ with $|I|=1$, and $2\binom{n/2-t-1}{2}2^{\frac{n}{2}-3}$ strings 
$\bbs\in S^{(t)}$ with $|I|=2$. 
To enumerate the strings in $S^{(t)}$ with $|I|\ge 3$, we need to calculate the size of $\mathcal{I}^{(t)}$. 
To that end, we let $i_0=t+1$ and $i_{l+1}=\frac{n}{2}$. Then for each $I=(i_1,\ldots,i_l)\in\mathcal{I}^{(t)}$, we can define $x_j=i_j-i_{j-1}$ where $j=1,\ldots,l+1$. Clearly, $x_1\ge 0,x_2,x_{l+1}\ge1$ and $x_j\ge 2$ for each $3\le j\le l$. Furthermore, we have $x_1+x_2+\cdots x_{l+1}=\frac{n}{2}-t-1$. Therefore, the size of $\mathcal{I}^{(t)}$ is equal to the number of integer solutions to the following equation
\begin{align}\label{eq_sizeofI}
   x_1+x_2+\cdots x_{l+1}=\frac{n}{2}-t-1 
\end{align}
under the condition that $x_1\ge 0,x_2\ge1,x_{l+1}\ge1$ and $x_j\ge 2$ for each $3\le j\le l$.

\begin{lemma}\label{lem_numofsolution}
  For given a given positive integer $N$, the number of integer solutions to the following equation
  \begin{equation*}
    y_1+\cdots +y_k=N
  \end{equation*}
  under the condition that $y_i\ge 0$ for all $i=1,\ldots,k$, is $\binom{N+k-1}{N}$. More generally, the number of integer solutions to the above equation under the condition that $x_i\ge p_i$ for all $i=1,\ldots,k$, is $\binom{N+k-\sum_{i=1}^{k}p_i-1}{k-1}$, where $p_1,\ldots,p_k$ are nonnegative integers.
\end{lemma}
\begin{IEEEproof}
  The first part is well known and appears in \cite[Proposition 1.5]{Jukna2011}. 
  To prove the general conclusion, let $y_i^{\prime}=y_i-p_i$ for each $1\le i\le k$. 
  Then each $y_i^{\prime}$ is a nonnegative integer. 
  The general conclusion follows from the first part.
\end{IEEEproof}

We now conclude from \Cref{eq_sizeofI} and Lemma \ref{lem_numofsolution} that $\left|\mathcal{I}^{(t)}\right|=\binom{\frac{n}{2}-t+1-l}{l}$ and thus the size of $S^{(t)}$ is
\begin{align}\label{eq_sizeofSt}
    \left|S^{(t)}\right|&=\sum_{l=3}^{\frac{n}{2}-t+1}2^{\frac{n}{2}-l-1}\binom{\frac{n}{2}-t+1-l}{l}M_{l+1}+2^{\frac{n}{2}-1}+\left(\frac{n}{2}-t\right)2^{\frac{n}{2}-2}+2\binom{n/2-t-1}{2}2^{\frac{n}{2}-3}\notag\\
    &=\sum_{l=0}^{\frac{n}{2}-t+1}2^{\frac{n}{2}-l-1}\binom{\frac{n}{2}-t+1-l}{l}M_{l+1},
\end{align}
where the binomial coefficient $\binom{a}{b}$ is define to be $0$ if $a<b$, and the term $M_{l+1}$ follows from the fact that $s_1s_{i_1}\cdots s_{i_l}$ is a Catlan-Bertrand string for each $\bbs\in S^{(t)}$.
\begin{corollary}
Let $t\ge 2$ be a fixed integer and $n>2t+8$, then the size of $\mathcal{C}_A^{(t)}$ is
$$
\sum_{i=0}^{\frac{n}{2}-t}\binom{\frac{n}{2}-t}{i}2^{\frac{n}{2}-t-i}M_i+\sum_{l=0}^{\frac{n}{2}-t+1}2^{\frac{n}{2}-l-1}\binom{\frac{n}{2}-t+1-l}{l}M_{l+1}.
$$
\end{corollary}
\begin{IEEEproof}
   Notice that $\left|\mathcal{C}_A^{(t)}\right|=\left|S^{(t)}\right|+\left|\mathcal{S}_R^{(t)}(n)\right|$. Then the conclusion follows from \Cref{eq_srt} and \Cref{eq_sizeofSt}.
\end{IEEEproof}

Recall that $M_{l+1}=\binom{l}{\frac{l}{2}}$ if $l$ is even and $M_{l+1}=\frac{1}{2}\binom{l+1}{\frac{l+1}{2}}$ if $l$ is odd. 
Proposition 1 in \cite{wz3} implies
\begin{align}\label{eq_boundsofSt}
     \frac{2^{\frac{n}{2}-1}}{\sqrt{\pi\left(\left\lfloor\frac{n+3}{4}-\frac{t}{2}\right\rfloor+1\right)}}\sum_{l=0}^{\frac{n}{2}-t+1}\binom{\frac{n}{2}-t+1-l}{l}\le \left|S^{(t)}\right|\le \frac{2^{\frac{n}{2}-1}}{\sqrt{\pi/2}}\sum_{l=0}^{\frac{n}{2}-t+1}\binom{\frac{n}{2}-t+1-l}{l}
\end{align}
To further estimate the order of magnitude of $\left|S^{(t)}\right|$, we need the following lemma.
\begin{lemma}\label{lem_lucas}
For any nonnegative integer $m$, we have
$$
L_m\triangleq\sum_{k=0}^m\binom{m-k}{k}=F_{m+1},
$$
where $F_{m+1}=\frac{(1+\sqrt{5})^{m+1}-(1-\sqrt{5})^{m+1}}{2^{m+1}\sqrt{5}}$ is the $(m+1)$th Fibonacci number.
\end{lemma}
\begin{IEEEproof}
   Notice that $\binom{m-k}{k}=\binom{m-k-1}{k}+\binom{m-k-1}{k-1}$ when $k\ge 1$ and $m\ge k$. So when $m\ge 2$, we have
   \begin{align*}
       L_m&=1+\sum_{k=1}^m\binom{m-k}{k}=1+\sum_{k=1}^{\left\lfloor\frac{m}{2}\right\rfloor}\binom{m-k}{k}\\
       &=1+\sum_{k=1}^{\left\lfloor\frac{m}{2}\right\rfloor}\binom{m-k-1}{k}+\sum_{k=1}^{\left\lfloor\frac{m}{2}\right\rfloor}\binom{m-k-1}{k-1}\\
       &=\sum_{k=0}^{\left\lfloor\frac{m}{2}\right\rfloor}\binom{m-1-k}{k}+\sum_{k=0}^{\left\lfloor\frac{m}{2}\right\rfloor-1}\binom{m-2-k}{k}\\
       &=\sum_{k=0}^{m-1}\binom{m-1-k}{k}+\sum_{k=0}^{m-2}\binom{m-2-k}{k}=L_{m-1}+L_{m-2}.
   \end{align*}
 This implies that $L_m$ satisfies the recurrence relation of Fibonacci numbers. On the other hand, $L_0=1=F_1$ and $L_1=2=F_2$. Therefore, $L_m=F_{m+1}$ for all $m\ge 0$.
\end{IEEEproof}
\begin{corollary}\label{cor_boundsofSt}
The lower bound of $\left|S^{(t)}\right|$ is $\Omega\left(\frac{(2\alpha)^{\frac{n}{2}}}{\sqrt{n}}\right)$, and the upper bound of $\left|S^{(t)}\right|$ is $O\left((2\alpha)^{\frac{n}{2}}\right)$, , where $\alpha=\frac{1+\sqrt{5}}{2}$.
\end{corollary}
\begin{IEEEproof}
   Theorem 8.1 in \cite{Koshy2017} asserts that $F_{m+1}=\left\lfloor\frac{\alpha^{m+1}}{\sqrt{5}}+\frac{1}{2}\right\rfloor$ for all $m\ge0$. 
   Then the conclusion follows from \Cref{eq_boundsofSt} and Lemma \ref{lem_lucas}.
\end{IEEEproof}

Since $\alpha\approx 1.618<2$, \Cref{eq_srt} and Corollary \ref{cor_boundsofSt} imply that the order of magnitude of $\left|S^{(t)}\right|$ is strictly smaller than that of $\left|\mathcal{S}_R^{(t)}(n)\right|$. However, since $\alpha$ is relatively close to $2$, we still add many strings into $\mathcal{S}_R^{(t)}(n)$ to obtain $\mathcal{C}_A^{(t)}$.

\section{$r$-Length Limited Composition Multiset}
\label{sec:RLL}
In this section we suggest a new reconstruction model. 
Recall that in order to store data, two molecules with different mass are synthesized to represent the 0 and 1 symbols. 
These molecules are combined into a string to represent the data that the user wishes to store. 
In order to read the data, the synthetic polymers are read by tandem mass (MS/MS) spectrometer. 
As part of the reading process, the polymer molecule breaks in a position which is chosen at random, creating a prefix and a suffix. 
The breaking of the molecule is usually made by collisions, bombardment of the molecule with electrons or photons, or a chemical reaction (a molecule can also break down due to instability).  
It is reasonable to assume that longer molecule fragments tend to break more often than shorter fragments. 
For example, using electron bombardment, it is more likely that the electron will hit a longer polymer fragment than  hitting a short fragment. 
As a result, when the mass readings of the fragments are obtained, the composition of long fragments will not appear in those readings. 
Leaving us with a composition multiset of shorter strings.

Since our results in this section apply to general alphabets, we assume our alphabet is a $q$-ary alphabet $\Sigma_q=\{a_i~:~a_i=i,0\le i\le q-1\}$, where $q\ge 2$ is an integer. Let $\Sigma_q^{*}$ denote the set of all strings over $\Sigma_q$ of finite length. For a string $\bbs\in\Sigma_q^{*}$, we define the composition $c(\bbs)$ of $\bbs$ to be $c(\bbs)=a_0^{d_0}a_1^{d_1}\cdots a_{q-1}^{d_{q-1}}$, where $d_i$ is the number of appearances of the symbol $a_i$ in $\bbs$ and we omit $a_i^{d_i}$ if $d_i=0$. For example, let $\Sigma=\mathset{0,1,2}$, $\bbs=001112$ and $\bbt=02002$, then $c(\bbs)=0^21^32^1$ and $c(\bbt)=0^32^2$. For a length-$n$ string $\bbs$ and an integer $1\le l\le n$, the notation $C_l(\bbs)$ can be similarly defined as in \Cref{sec_delcomp}.

To model the phenomenon described as above, dubbed \textit{the $r$-length limited 
composition multiset model}, we assume that the composition multiset does not contain the composition of substrings longer than $r$. 
The mathematical model is defined next. 
\begin{definition} 
For a string $\bbs\in\Sigma_q^n$ and for a positive integer $r\leq n$, we denote by $C_{\le r}(\bbs)$ the multiset of compositions of all substrings of length 
at most $r$ in $\bbs$, i.e., 
\[C_{\le r}(\bbs)=\bigcup_{i=1}^r C_i(\bbs).\]
\end{definition}

In this section we initiate the study of $r$-length limited composition multiset model, 
and present a simple coding scheme for the case where $r=r(n)$ is a linear function of the length of the strings $n$. 
To that end, we define the following relation on strings.
For $\bbu,\bbv\in\Sigma_q^n$, we say that $\bbu,\bbv$ are cyclic-equivalent, 
denoted $\bbu\overset{c}{\sim} \bbv$, if $\bbu$ is a cyclic shift of $\bbv$. It is immediate that $\overset{c}{\sim}$ is an equivalence relation. 
A string $\bbu$ is called \textit{minimal} if it is the minimal (with respect to the lexicographic order) string among all its equivalent strings. 
A string $\bbu\in\Sigma_2^n$ is called \textit{periodic} if there exists a string $\bbv$ such 
that $\bbu=\bbv^t$ with $t\geq 2$, where $\bbv^t$ denotes the consecutive concatenation of $\bbv$ with itself $t$ times. 
A string is called \textit{aperiodic} if it is not periodic.
\begin{definition}
A string $\bbw\in\Sigma_q^*$ is called a \textit{Lyndon string} if $\bbw$ is minimal and aperiodic. 
We denote by $\cL_r(q)$ the set of all Lyndon words of length $r$ over $\Sigma_q$.
\end{definition} 
Lyndon words were first introduced in \cite{Lyn1954,Lyn1955} and were used to study free abelian groups. They have been found useful in various topics such as differential equations, cryptography, and coding theory \cite{Li1997,Sub1992,Eli2021}. 

For a Lyndon word $\bbl\in\cL_r(q)$, denote by $m$ the maximal number of 
consecutive $a_q$  
symbols that appear at the end of $\bbl$, i.e., $\bbl=l_1l_2\dots l_{r-m}a_q^m$ 
where $m\geq 0$ is maximal and $l_{r-m}< a_q$. 
We construct our code $\cC_r$ as follows 
\[ \cC_r=\mathset{\bbl a_q^{r-m}  ~:~ \bbl\in\cL_r(q)}.\] 
So a codeword $\bbs\in\cC_r$ is a Lyndon string of length $r$ followed by a string of repeated $a_q$ such that $\bbs$ ends with $r$ consecutive $a_q$ symbols. 
The strings in $\cC_r$ have variable length.
The shortest codeword has $r+1$ symbols (obtained by Lyndon words of the form 
$a_i a_q^{r-1}$ for $i<q-1$, and then $\bbs=a_ia_q^r$), 
and the longest codeword has $2r$ symbols if $q>2$ and $2r-1$ symbols if the alphabet contains two symbols (obtained by the Lyndon words $a_i\dots a_ia_j$ with $i<j<q$, 
concatenated with $a_q^r$ to generate $\bbs=a_i^{r-1} a_j a_q^r$). 
If length-$n$ codewords are in need, where $n\geq 2r$, such a set $\cC_r(n)$ can be obtained by padding every codeword in $\cC_r$ with $a_q$ until it reaches length $n$.

The main goal of this section is to prove the following corollary which will be readily obtained from the discussion in this section. 
\begin{corollary}
Let $\Sigma_q$ be an alphabet of size $q$ and consider the $r$-length limited composition multiset model with a fixed $r$. 
Every codeword $\bbs\in\cC_r$ is uniquely reconstructible given $C_{\le r}(\bbs)$ 
and the code rate is 
\[R=\frac{\log_q \parenv{|\cC_r|}}{\text{Avg. length of }\bbs\in \cC_r}\geq \frac{1}{2}\parenv{1+o(1)}.\]
\end{corollary}

In order to generate the set $\cL_r(q)$ we can use the efficient algorithm given in \cite{Duv1983,Duv1988} for finding Lyndon words of length at most $r$, according to their lexicographic order. 
The algorithm is presented here for completeness. 
\begin{algorithm}
	\caption{Algorithm that generates next Lyndon words (up to length $r$)}
	\label{alg:1}
	\begin{algorithmic}[1]
		\vspace{.1ex}
		\Require A Lyndon word $\bbw\in\Sigma_q^*$. 
		\Ensure The next Lyndon word $\bbu$ according to the lexicographic order. 
		\State Let $\bbs=(\bbw^r)_1^r$, i.e., concatenate $\bbw$ to itself to form a string $\bbs\in\Sigma_q^r$ with $\bbs_i=\bbw_{i\bmod |\bbw|}$. 
        \While{ last symbol of $\bbs$ is the maximal symbol in lexicographic order}
		\State Remove the last symbol from $\bbs$, i.e., $\bbs=\bbs_1^{|\bbs|-1}$.
		\EndWhile
		\State Replace the last symbol in $\bbs$ by its successor according to the Lexicographic order to generate $\bbu$. 
		\State Return $\bbu$
			\end{algorithmic}
\end{algorithm}

\begin{example}
Let $\Sigma_q=\mathset{0,1,2,3}$ with the usual ordering and let $\bbw=003103$ be a Lyndon word 
of length $|\bbw|=6$. 
Applying Algorithm \ref{alg:1} with $r=9$ we first generate $\bbs=003103003$. 
Next, we eliminate the last symbol since it is the maximal symbol in our alphabet, to obtain $\bbs=00310300$. Lastly, we replace the last symbol $0$ with its successor $1$ to obtain 
$\bbu=00310301$. 

Notice that the algorithm may also provide a string which is shorter than $r$. 
Consider the Lyndon word $\bbw=00000333$. 
Eliminating the maximal symbol in Step 3 of the algorithm yields $\bbs=000000$ (as the output at Step 4) and the next Lyndon word obtained is $\bbu=000001$ which is of length $6$.
\end{example} 

Instead of showing that every codeword $\bbs\in\cC_r$ can be uniquely reconstructed from $C_{\le r}(\bbs)$, we show that $\bbs$ can be uniquely reconstructed from the multiset $C_r(\bbs)$ of compositions of substring of length $r$. 
Indeed, notice that there are no Lyndon words of length $r$ with $r$ appearances of $a_q$. 
This implies that $a_q^r$ appears exactly once in $C_r(\bbs)$. 
From that point onward, we can construct $\bbs$ backwards by considering compositions containing a decreasing number of $a_q$s. 

The number of length-$r$ Lyndon words over an alphabet $\Sigma_q$ is given by Witt's formula \cite{Cra2012}  
\[|\cL_r(q)|=\frac{1}{r}\sum_{d|r}\mu(d)q^{r/d}\] 
where $\mu$ is the M\"{o}ebius function. 

Since $\mu(\cdot)\in\mathset{0,1,-1}$, we can bound the size $|\cL_r(q)|$. 
Notice that $\mu(1)=1$ which implies that 
\begin{align*}
|\cL_r(q)|&=\frac{1}{r}q^r+\frac{1}{r}\sum_{\substack{d|r \\ d>1}}\mu(d)q^{r/d} \\ 
&\leq \frac{1}{r}q^r+q^{r/2}
\end{align*}
and similarly, 
\[|\cL_r(q)|\geq \frac{1}{r}q^r-q^{r/2}.\]
Thus, the number of Lyndon words over an alphabet of size $q$ is $|\cL_r(q)|=\frac{q^r}{r}\parenv{1+o(1)}$.

The length of a codeword $\bbs\in\cC_r$ is bounded above by $2r$. 
Thus, we obtain that 
\begin{align*}
    R&\geq \frac{\log_q |\cL_r(q)|}{2r}\\
    &=\frac{r-\log_q (r) +\log_q(1+o(1)}{2r}\\
    &=\frac{1}{2}-\frac{\log_q(r)}{r}+ o(1)\\
    &= \frac{1}{2}(1+o(1)).
\end{align*}

\begin{remark}
A similar encoding process can be used to reconstruct a string $\bbs\in\cC_r(n)$ when $n\geq 2r$. Thus, it is possible to use $\cC_r(n)$ as a code of length $n$. 
Applying similar calculations with $n$ instead of $2r$, we obtain code rate $R=r/n$. 
Thus, when $r=r(n)$ is a linear function of $n$, such as $r(n)=n/\alpha$ for some $\alpha\in \N$, we obtain a rate of $1/\alpha$. If $r(n)=o(n)$ the code rate is $0$.
\end{remark}

\section{Conclusion}\label{sec_conclusion}
In this paper, we firstly solved the problem on whether a string can be uniquely reconstructed up to reversal or not, when only given its prefix-suffix composition multiset:
\begin{itemize}
  \item we completely characterized their structures;
  \item we give an explicit formula of their numbers;
  \item for any given string $\bbs$, we gave a method to find all the strings that share a common prefix-suffix composition multiset with $\bbs$, and we gave a formula of the number of such strings.
\end{itemize}

As applications of these results, we gave a construction of composition codes of maximum size. But we do not know whether or not there is an efficient reconstruction algorithm to reconstruct an arbitrary string in this kind of code. This will be an interesting problem for the future research. As a compromise, we constructed a subcode $\mathcal{E}_1(n)$ of some composition code of maximum size. For this subcode, there does exist a reconstruction algorithm. It is shown that the redundancy of this subcode is less that $\frac{1}{2}\log_2(n)+3+\frac{1}{2}\log_2(\pi)$.

In addition, we studied two error models of composition errors. For each error model, we modified $\mathcal{E}_1(n)$ to get a composition code that can correct the studied type of error; each code has asymptotic rate $1$.

The authors of \cite{wz2} focused on the problem of uniquely reconstructing a set of strings from the union of their prefix-suffix composition multisets. We believe that our results in this paper might be a helpful step towards fully understanding this problem. It is also interesting to extend our results to this situation.
Moreover, in either this paper or \cite{wz2}, the constructed codes uses the idea that we can distinguish the prefix composition multiset from the suffix composition multiset. 
An interesting question is whether it is possible to construct codes without this property, such that a set of strings can be uniquely reconstructed from the union of their prefix-suffix composition multiset. We leave this problem for a future endeavour. 

Secondly, we initiated the problem of reconstructing a string from its composition multiset under the condition that at most a constant number of its compositions are lost. We solved this problem for a special case when the error is asymmetric. It is interesting to introduce new ideas to solve this problem for more parameters and even without the "asymmetric" assumption.

Lastly, we presented the problem of $r$-length limited composition multiset reconstruction, 
and provided a simple coding scheme that obtains a rate of $1/2$. 
This problem gives rise to many interesting problems. 
For example, is it possible to improve the code presented in this problem for the case $r=r(n)$ is a linear function of $n$? What about codes for a constant $r$ or for a different functions $r(n)$? 
Is it possible to find bounds on the capacity of such codes? 
Also, some generalizations of the model itself are of interest. For example, consider the case in which compositions of substrings of lengths between $r_1$ and $r_2$ are obtained, 
Consider the appearances of errors, or even considering a scattered noise model in which the compositions are obtained according to a probability which is a function of their length. 
As an example, consider the case that all the compositions up to length $r_1$ are obtained, and the compositions of strings of length $r_1+j$ are obtained with a probability of $1/2^{j-1}$.

\bibliographystyle{IEEEtran}
\bibliography{ref}

\begin{thebibliography}{10}
\providecommand{\url}[1]{#1}
\csname url@samestyle\endcsname
\providecommand{\newblock}{\relax}
\providecommand{\bibinfo}[2]{#2}
\providecommand{\BIBentrySTDinterwordspacing}{\spaceskip=0pt\relax}
\providecommand{\BIBentryALTinterwordstretchfactor}{4}
\providecommand{\BIBentryALTinterwordspacing}{\spaceskip=\fontdimen2\font plus
\BIBentryALTinterwordstretchfactor\fontdimen3\font minus
  \fontdimen4\font\relax}
\providecommand{\BIBforeignlanguage}[2]{{%
\expandafter\ifx\csname l@#1\endcsname\relax
\typeout{** WARNING: IEEEtran.bst: No hyphenation pattern has been}%
\typeout{** loaded for the language `#1'. Using the pattern for}%
\typeout{** the default language instead.}%
\else
\language=\csname l@#1\endcsname
\fi
#2}}
\providecommand{\BIBdecl}{\relax}
\BIBdecl

\bibitem{IDC2018}
\BIBentryALTinterwordspacing
J.~G. David~Reinsel and J.~Rydning, ``The {D}igitization of the {W}orld from
  {E}dge to {C}ore,'' in \emph{IDC Report}, 2018. [Online]. Available:
  \url{https://www.seagate.com/files/www-content/our-story/trends/files/idc-seagate-dataage-whitepaper.pdf}
\BIBentrySTDinterwordspacing

\bibitem{science2012}
G.~M. {Church}, Y.~{Gao}, and S.~{Kosuri}, ``Next-{G}eneration {D}igital
  {I}nformation {S}torage in {DNA},'' \emph{Science}, vol. 337, no. 6102, pp.
  1628--1628, Sep. 2012.

\bibitem{nature2013}
N.~{Goldman}, P.~{Bertone}, S.~{Chen}, C.~{Dessimoz}, E.~M. {LeProust},
  B.~{Sipos}, and E.~{Birney}, ``Towards {P}ractical, {H}igh-capacity,
  {L}ow-{M}aintenance {I}nformation {S}torage in {S}ynthesized {DNA},''
  \emph{Nature}, vol. 494, no. 7435, pp. 77--80, Feb. 2013.

\bibitem{wz4}
R.~N. Grass, R.~Heckel, M.~Puddu, D.~Paunescu, and W.~J. Stark, ``Robust
  {C}hemical {P}reservation of {D}igital {I}nformation on {DNA} in {S}ilica
  with {E}rror-{C}orrecting {C}odes,'' \emph{Angewandte Chemie International
  Edition}, vol.~54, no.~8, pp. 2552--2555, Feb. 2015.

\bibitem{wz5}
S.~M. H.~T. Yazdi, Y.~Yuan, J.~Ma, H.~Zhao, and O.~Milenkovic, ``A
  {R}ewritable, {R}andom-{A}ccess {DNA}-{B}ased {S}torage {S}ystem,''
  \emph{Scientific Reports}, vol.~5, p. 14318, Sep. 2015.

\bibitem{wz6}
V.~Zhirnov, R.~M. Zadegan, G.~S. Sandhu, G.~M. Church, and W.~L. Hughes,
  ``Nucleic acid memory,'' \emph{Nature Materials}, vol.~15, no.~4, pp.
  366--370, Apr. 2016.

\bibitem{wz7}
A.~A. Ouahabi, J.-A. Amalian, L.~Charles, and J.-F. Lutz, ``Mass spectrometry
  sequencing of long digital polymers facilitated by programmed inter-byte
  fragmentation,'' \emph{Nature Communications}, vol.~8, p. 967, Oct. 2017.

\bibitem{wz8}
S.~M. H.~T. Yazdi, R.~Gabrys, and O.~Milenkovic, ``Portable and {E}rror-{F}ree
  {D}na-{B}ased {D}ata {S}torage,'' \emph{Scientific Reports}, vol.~7, p. 5011,
  Jul. 2017.

\bibitem{wz9}
M.~G. T.~A. Rutten, F.~W. Vaandrager, J.~A. A.~W. Elemans, and R.~J.~M. Nolte,
  ``Encoding information into polymers,'' \emph{Nature Reviews Chemistry},
  vol.~2, no.~11, pp. 365--381, Nov. 2018.

\bibitem{wz10}
R.~Heckel, G.~Mikutis, and R.~N. Grass, ``A {C}haracterization of the {DNA}
  {D}ata {S}torage {C}hannel,'' \emph{Scientific Reports}, vol.~9, p. 9663,
  Jul. 2019.

\bibitem{wz11}
C.~N. Takahashi, B.~H. Nguyen, K.~Strauss, and L.~Ceze, ``Demonstration of
  {E}nd-to-{E}nd {A}utomation of {DNA} {D}ata {S}torage,'' \emph{Scientific
  Reports}, vol.~9, p. 4998, Mar. 2019.

\bibitem{wz12}
S.~K. Tabatabaei, B.~Wang, N.~B. Murali~Athreya, B.~Enghiad, A.~G. Hernandez,
  J.-P. Leburton, D.~Soloveichik, H.~Zhao, and O.~Milenkovic, ``{DNA} {P}unch
  {C}ards: {E}ncoding {D}ata on {N}ative {DNA} {S}equences via {T}opological
  {M}odifications,'' \emph{bioRxiv}, p. 672394, 2019.

\bibitem{wz13}
S.~K. Tabatabaei, B.~Wang, N.~B.~M. Athreya, B.~Enghiad, A.~G. Hernandez, C.~J.
  Fields, J.-P. Leburton, D.~Soloveichik, H.~Zhao, and O.~Milenkovic, ``{DNA}
  punch cards for storing data on native {DNA} sequences via enzymatic
  nicking,'' \emph{Nature Communications}, vol.~11, p. 1742, Apr. 2020.

\bibitem{wz14}
K.~Matange, J.~M. Tuck, and A.~J. Keung, ``{DNA} stability: a central design
  consideration for {DNA} data storage systems,'' \emph{Nature Communications},
  vol.~12, p. 1358, Mar. 2021.

\bibitem{wz15}
K.~Launay, J.-A. Amalian, E.~Laurent, L.~Oswald, A.~A. Ouahabi, A.~Burel,
  F.~Dufour, C.~Carapito, J.-L. Clement, J.-F. Lutz, L.~Charles, and D.~Gigmes,
  ``Precise {A}lkoxyamine {D}esign to {E}nable {A}utomated {T}andem {M}ass
  {S}pectrometry {S}equencing of {D}igital {P}oly(phosphodiester)s,''
  \emph{Angewandte Chemie International Edition}, vol.~60, no.~2, pp. 917--926,
  Jan. 2021.

\bibitem{wz16}
C.~Pan, S.~K. Tabatabaei, S.~M. H.~T. Yazdi, A.~G. Hernandez, C.~M. Schroeder,
  and O.~Milenkovic, ``{DNA} stability: a central design consideration for
  {DNA} data storage systems,'' \emph{Nature Communications}, vol.~13, p. 2984,
  May 2022.

\bibitem{wz1}
J.~Acharya, H.~Das, O.~Milenkovic, A.~Orlitsky, and S.~Pan, ``String
  {R}econstruction from {S}ubstring {C}ompositions,'' \emph{SIAM Journal on
  Discrete Mathematics}, vol.~29, no.~3, pp. 1340--1371, 2015.

\bibitem{wz3}
\BIBentryALTinterwordspacing
S.~Pattabiraman, R.~Gabrys, and O.~Milenkovic, ``Coding for {P}olymer-{B}ased
  {D}ata {S}torage,'' \emph{arXiv: 2003.02121v2}, 2021. [Online]. Available:
  \url{http://arxiv.org/abs/2003.02121v2}
\BIBentrySTDinterwordspacing

\bibitem{wz17}
\BIBentryALTinterwordspacing
A.~Banerjee, A.~Wachter-Zeh, and E.~Yaakobi, ``Insertion and {D}eletion
  {C}orrection in {P}olymer-based {D}ata {S}torage,'' \emph{arXiv:
  2201.08612v2}, 2022. [Online]. Available:
  \url{https://arxiv.org/abs/2201.08612v2}
\BIBentrySTDinterwordspacing

\bibitem{wz18}
\BIBentryALTinterwordspacing
U.~Gupta and H.~Mahdavifar, ``A {N}ew {A}lgebraic {A}pproach for {S}tring
  {R}econstruction from {S}ubstring {C}ompositions,''
  \emph{arXiv:2201.09955v1}, 2022. [Online]. Available:
  \url{https://arxiv.org/abs/2201.09955v1}
\BIBentrySTDinterwordspacing

\bibitem{wz2}
\BIBentryALTinterwordspacing
R.~Gabrys, S.~Pattabiraman, and O.~Milenkovic, ``Reconstruction of {S}ets of
  {S}trings from {P}refix/{S}uffix {C}ompositions,'' \emph{arXiv:
  2110.02352v1}, 2021. [Online]. Available:
  \url{http://arxiv.org/abs/2110.02352v1}
\BIBentrySTDinterwordspacing

\bibitem{gtm290}
\"{O}mer E\u{g}ecio\u{g}lu and A.~M. Garsia, \emph{Lessons in Enumerative
  Combinatorics}, 1st~ed., ser. Graduate Texts in Mathematics.\hskip 1em plus
  0.5em minus 0.4em\relax Springer Cham, 2021, vol. 290.

\bibitem{huffman_pless_2003}
W.~C. Huffman and V.~Pless, \emph{Fundamentals of Error-Correcting
  Codes}.\hskip 1em plus 0.5em minus 0.4em\relax Cambridge University Press,
  2003.

\bibitem{Jukna2011}
S.~Jukna, \emph{Extremal Combinatorics}, 2nd~ed., ser. Texts in Theoretical
  Computer Science. An EATCS Series.\hskip 1em plus 0.5em minus 0.4em\relax
  Springer Berlin, Heidelberg, 2011.

\bibitem{Koshy2017}
T.~Koshy, \emph{Fibonacci and Lucas Numbers with Applications (Volume 1)},
  2nd~ed., ser. Pure and Applied Mathematics: A Wiley Series of Texts,
  Monographs and Tracts.\hskip 1em plus 0.5em minus 0.4em\relax John Wiley \&
  Sons, Ltd, 2017.

\bibitem{Lyn1954}
R.~C. Lyndon, ``On {B}urnside’s {P}roblem,'' \emph{Transactions of the
  American Mathematical Society}, vol.~77, no.~2, pp. 202--215, Sep. 1954.

\bibitem{Lyn1955}
------, ``On {B}urnside's {P}roblem. {II},'' \emph{Transactions of the American
  Mathematical Society}, vol.~78, no.~2, pp. 329--332, Mar. 1955.

\bibitem{Li1997}
C.~W. Li and W.~Q. Liu, ``Approximation of multiple stochastic integrals and
  its application to stochastic differential equations,'' \emph{Nonlinear
  Analysis: Theory, Methods \& Applications}, vol.~30, no.~2, pp. 697--708,
  Dec. 1997.

\bibitem{Sub1992}
K.~G. Subramanian, R.~Siromoney, and L.~Mathew, ``Lyndon trees,''
  \emph{Theoretical computer science}, vol. 106, no.~2, pp. 373--383, Dec.
  1992.

\bibitem{Eli2021}
O.~Elishco, R.~Gabrys, E.~Yaakobi, and M.~M{\'e}dard, ``Repeat-{F}ree
  {C}odes,'' \emph{IEEE Transactions on Information Theory}, vol.~67, no.~9,
  pp. 5749--5764, Jun. 2021.

\bibitem{Duv1983}
J.~P. Duval, ``Factorizing words over an ordered alphabet,'' \emph{Journal of
  Algorithms}, vol.~4, no.~4, pp. 363--381, Dec. 1983.

\bibitem{Duv1988}
------, ``G{\'e}n{\'e}ration of a section of conjugation classes and {L}yndon
  word tree of length born{\'e}e,'' \emph{Theoretical computer science},
  vol.~60, no.~3, pp. 255--283, Dec. 1988.

\bibitem{Cra2012}
H.~Crapo and D.~Senato, \emph{Algebraic Combinatorics and Computer Science: A
  Tribute to Gian-Carlo Rota}.\hskip 1em plus 0.5em minus 0.4em\relax Springer
  Science \& Business Media, 2012.

\end{thebibliography}

\end{document}